\documentclass[
  journal=pasa,
  manuscript=research-paper,
  year=2024,
  volume=xx,
]{cup-journal}

\usepackage{graphicx}	
\usepackage{amsmath}	
\usepackage{amssymb}	
\usepackage{bm}
\usepackage[tracking=smallcaps]{microtype}
\SetTracking[no ligatures = f]{encoding = *,shape = sc}{0}
\usepackage{booktabs}

\usepackage{natbib} 
\usepackage{verbatim}
\usepackage{multirow}
\usepackage{subfigure}

\usepackage[T1]{fontenc}
\usepackage[utf8]{inputenc}

\usepackage{footnote}

\makesavenoteenv{tabular}
\makesavenoteenv{table}
\usepackage{supertabular}
\usepackage{threeparttable}
\usepackage{longtable}
\usepackage{rotating}
\usepackage{lscape}

\usepackage{lineno,hyperref}
\modulolinenumbers[5]
\hypersetup{colorlinks,citecolor=blue,linkcolor=blue,urlcolor=blue}

\usepackage[]{algorithm2e}
\usepackage{dsfont}
\usepackage{siunitx}

\usepackage{float}

\definecolor{babypink}{rgb}{0.96, 0.76, 0.76}
\definecolor{beaublue}{rgb}{0.74, 0.83, 0.9}
\definecolor{mediumturquoise}{rgb}{0.28, 0.82, 0.8}
\definecolor{mossgreen}{rgb}{0.68, 0.87, 0.68}
\definecolor{mustard}{rgb}{1.0, 0.86, 0.35}
\definecolor{olivine}{rgb}{0.6, 0.73, 0.45}
\definecolor{orchid}{rgb}{0.85, 0.44, 0.84}
\definecolor{palecerulean}{rgb}{0.61, 0.77, 0.89}
\definecolor{palegold}{rgb}{0.9, 0.75, 0.54}
\definecolor{paleplum}{rgb}{0.8, 0.6, 0.8}
\definecolor{palespringbud}{rgb}{0.93, 0.92, 0.74}
\definecolor{pastelgray}{rgb}{0.81, 0.81, 0.77}
\definecolor{pastelviolet}{rgb}{0.8, 0.6, 0.79}
\definecolor{pastelred}{rgb}{1.0, 0.41, 0.38}
\definecolor{pearl}{rgb}{0.94, 0.92, 0.84}
\definecolor{pistachio}{rgb}{0.58, 0.77, 0.45}	
\definecolor{teal}{rgb}{0.0, 0.5, 0.5}
\definecolor{tiffanyblue}{rgb}{0.04, 0.73, 0.71}
\definecolor{turquoise}{rgb}{0.19, 0.84, 0.78}
\definecolor{verdigris}{rgb}{0.26, 0.7, 0.68}
	
\definecolor{lightgray}{gray}{0.9}

\title{Self-supervised contrastive learning of radio data for source detection, classification and peculiar object discovery}

\author{S. Riggi}
\affiliation{INAF - Osservatorio Astrofisico di Catania, Via Santa Sofia 78, 95123 Catania, Italy}
\email[S. Riggi]{simone.riggi@inaf.it}

\author{T. Cecconello}
\affiliation{INAF - Osservatorio Astrofisico di Catania, Via Santa Sofia 78, 95123 Catania, Italy}
\alsoaffiliation{Department of Electrical, Electronic and Computer Engineering, University of Catania, Viale Andrea Doria 6, 95125 Catania, Italy}

\author{S. Palazzo}
\affiliation{Department of Electrical, Electronic and Computer Engineering, University of Catania, Viale Andrea Doria 6, 95125 Catania, Italy}

\author{A.M. Hopkins}
\affiliation{School of Mathematical and Physical Sciences, 12 Wally’s Walk, Macquarie University, NSW 2109, Australia}

\author{N. Gupta}
\affiliation{CSIRO Space \& Astronomy, PO Box 1130, Bentley WA 6102, Australia}

\author{C. Bordiu}
\affiliation{INAF - Osservatorio Astrofisico di Catania, Via Santa Sofia 78, 95123 Catania, Italy}

\author{A. Ingallinera}
\affiliation{INAF - Osservatorio Astrofisico di Catania, Via Santa Sofia 78, 95123 Catania, Italy}

\author{C. Buemi}
\affiliation{INAF - Osservatorio Astrofisico di Catania, Via Santa Sofia 78, 95123 Catania, Italy}

\author{F. Bufano}
\affiliation{INAF - Osservatorio Astrofisico di Catania, Via Santa Sofia 78, 95123 Catania, Italy}

\author{F. Cavallaro}
\affiliation{INAF - Osservatorio Astrofisico di Catania, Via Santa Sofia 78, 95123 Catania, Italy}

\author{M.D. Filipovi\'c}
\affiliation{Western Sydney University, Locked Bag 1797, Penrith South DC, NSW 2751, Australia}%

\author{P. Leto} 
\affiliation{INAF - Osservatorio Astrofisico di Catania, Via Santa Sofia 78, 95123 Catania, Italy}

\author{S. Loru}
\affiliation{INAF - Osservatorio Astrofisico di Catania, Via Santa Sofia 78, 95123 Catania, Italy}

\author{A.C. Ruggeri}
\affiliation{INAF - Osservatorio Astrofisico di Catania, Via Santa Sofia 78, 95123 Catania, Italy}

\author{C. Trigilio}
\affiliation{INAF - Osservatorio Astrofisico di Catania, Via Santa Sofia 78, 95123 Catania, Italy}

\author{G. Umana}
\affiliation{INAF - Osservatorio Astrofisico di Catania, Via Santa Sofia 78, 95123 Catania, Italy}

\author{F. Vitello}
\affiliation{INAF - Osservatorio Astrofisico di Catania, Via Santa Sofia 78, 95123 Catania, Italy}

\keywords{Radio sources, Radio source catalogs, Astronomy image processing, Deep learning, Classification, Outlier detection}

\begin{document}
\sloppy

\begin{abstract}
New advancements in radio data post-processing are underway within the SKA precursor community, aiming to facilitate the extraction of scientific results from survey images through a semi-automated approach. Several of these developments leverage deep learning (DL) methodologies for diverse tasks, including source detection, object or morphology classification, and anomaly detection. Despite substantial progress, the full potential of these methods often remains untapped due to challenges associated with training large supervised models, particularly in the presence of small and class-unbalanced labelled datasets.\\Self-supervised learning has recently established itself as a powerful methodology to deal with some of the aforementioned challenges, by directly learning a lower-dimensional representation from large samples of unlabelled data. The resulting model and data representation can then be used for data inspection and various downstream tasks if a small subset of labelled data is available.\\
In this work, we explored contrastive learning methods to learn suitable radio data representation from unlabelled images taken from the ASKAP EMU and SARAO MeerKAT GPS surveys. We evaluated trained models and the obtained data representation over smaller labelled datasets, also taken from different radio surveys, in selected analysis tasks: source detection and classification, and search for objects with peculiar morphology. For all explored downstream tasks, we reported and discussed the benefits brought by self-supervised foundational models built on radio data. 
\end{abstract}


\section{Introduction}
\label{sec:intro}
Radio astronomy stands at the threshold of a transformative era, marked by the advent of large sky surveys carried out with instruments such as the Square Kilometre Array (SKA) \citep{SKADesignDoc} and its precursor telescopes. As the field enters this golden age, the immense volumes of observational data generated pose unprecedented challenges and opportunities. For example, the Evolutionary Map of the Universe (EMU) \citep{Norris2011} of the Australian SKA Pathfinder (ASKAP, \citealt{Johnston2008,ASKAPSystemDesign}) started in 2022 to survey $\sim$75\% of the sky at 940 MHz with an angular resolution of $\sim$15" and a noise rms of $\sim$15 $\mu$Jy/beam. The EMU source cataloguing process will require an unprecedented degree of automation and knowledge extraction, as the expected number of detectable sources is $\sim$50 millions. So will be for other precursors and future SKA observations. The sheer scale and complexity of these datasets demand innovative approaches to shorten the time needed to deliver scientific results or groundbreaking discoveries.\\In this context, machine learning (ML) emerges as a powerful tool for unlocking the full potential of radio astronomy data, offering solutions to complex tasks that are often beyond the reach of conventional methods in multiple areas, including source extraction, classification (e.g. morphological or astrophysical type) and discovery of anomalous/unexpected objects. 
\begin{figure*}[htb]
\centering%
\includegraphics[scale=0.68]{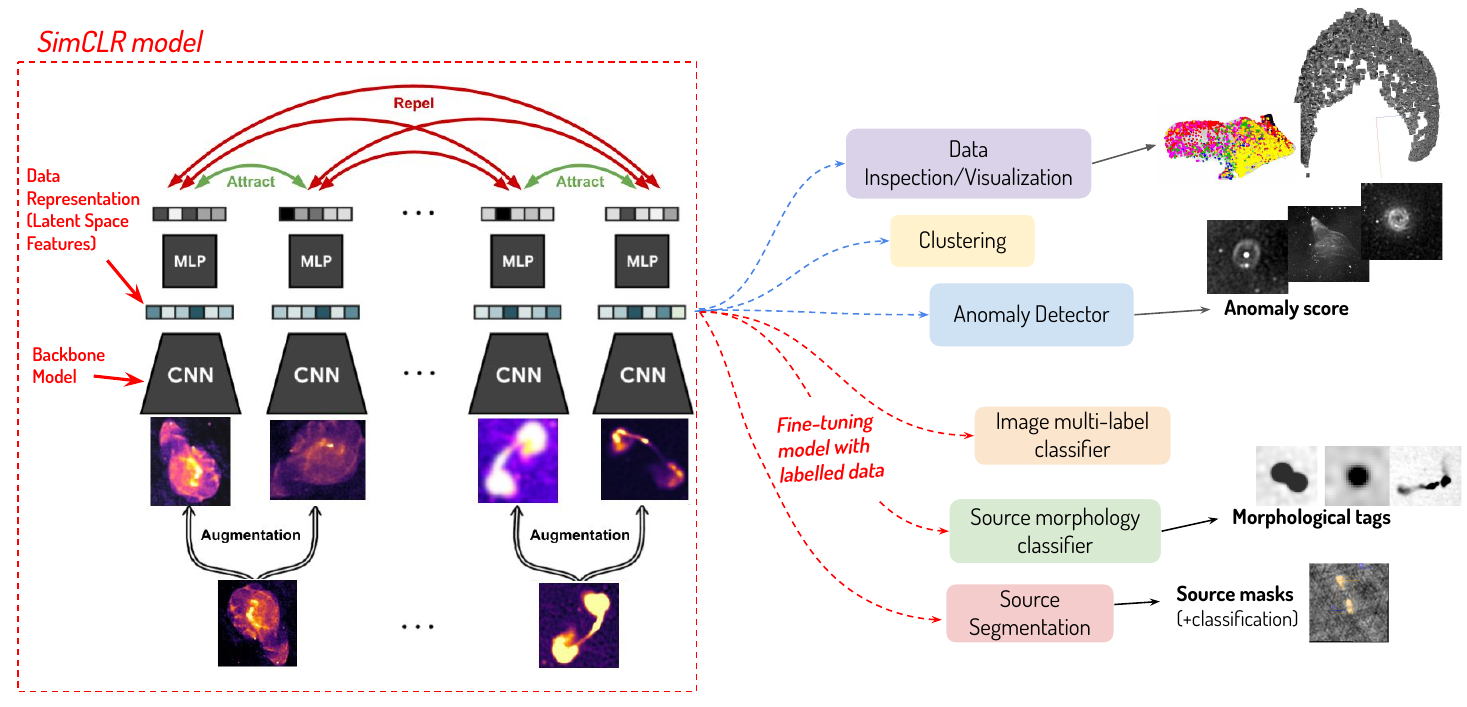}%
\caption{Schema of self-supervised learning for radio data analysis.}%
\label{fig:ssc_schema}
\end{figure*}
Most existing contributions focused on galaxy morphology classification for extragalactic science cases employing either supervised learning (SL), e.g. with convolutional neural networks (CNNs) \citep{Aniyan2017,Lukic2018,Wu2019,Lao2023} or Vision Transformers (ViTs) \citep{Gupta2024}, weakly-supervised learning \citep{Gupta2023}, semi-supervised learning \citep{Slijepcevic2022}, or unsupervised learning, e.g. Self-Organizing Maps (SOMs) \citep{Galvin2020,Mostert2021,Gupta2022} or t-distributed stochastic neighbour embedding \citep{Pennock2022}.\\Despite substantial progress, the full potential of supervised approaches often remains untapped due to the scarcity of large and high-quality annotated radio image datasets, crucial for training very deep models. The human effort required to produce them is in fact unsustainable. Citizen science projects, launched within different precursor surveys on the Zooniverse platform\footnote{\url{https://www.zooniverse.org/projects/chrismrp/radio-galaxy-zoo-lofar}}\footnote{\url{https://www.zooniverse.org/projects/hongming-tang/radio-galaxy-zoo-emu}} and building on the previous Radio Galaxy Zoo project \citep{RGZ}, will partially alleviate this need, at the cost of potentially introducing errors and biases in the cumulative dataset. As a result, existing labelled radio datasets are typically very limited in size, class-unbalanced, and adopt a diverse or ambiguous labelling schema, usually depending on the particular domain of application. Several applications produced so far for either radio source classification or source detection scopes, have therefore resorted to fine-tune models that were previously pre-trained on large annotated collections of non-astronomical data, such as the \textit{ImageNet} \citep{ImageNet} or \textit{COCO} \citep{COCO} datasets, that may not well capture all distinctive features of radio observations. On the other hand, completely unsupervised approaches are not very effective when directly dealing with highly dimensional image data, typically requiring previous feature extraction and dimensionality reduction steps to be applied. Currently, employed methods based on SOMs typically enforce an apriori discrete static data organization that do not well support extension to new tasks. These limitations necessitate exploring alternative methodologies.\\Representation learning \citep{Bengio2013}, and in particular self-supervised learning (SSL) \citep{Liu2023}, has recently emerged as a promising avenue to address these issues, by directly learning (pretext task), without any supervision, a lower-dimensionality representation (i.e. the latent space) from large samples of unlabelled data. The resulting model and data representation can then be used for data inspection and generalized for various applications (downstream tasks), e.g. classification, object detection, etc, if a small subset of labelled data is available.
Previous works in the radio domain are based on convolutional autoencoders (CAE) generative methods, which learns a latent space by minimizing a loss between input data and data reconstructed through an encoder-decoder network. For example, \citealt{Ralph2019} developed a pipeline for unsupervised source morphology studies based on SOMs and k-mean clustering algorithm, employing CAEs to extract features from the Radio Galaxy Zoo (RGZ) \citep{RGZ} images. \cite{BordiuSNR} employed CAEs to extract features from combined radio and infrared images of known Galactic supernova remnants (SNRs) to search for possible multiwavelength patterns.\\Contrastive learning approaches, on the other hand, employ siamese or teacher-student network architectures, minimizing the similarity between augmented versions of the input data, eventually in contrast to negative samples. They were reported to obtain superior performance on natural images in classification tasks (e.g. rivalling supervised methods), quality of representation learnt, computation efficiency, and robustness to noise. Recently, \cite{Slijepcevic2024} trained BYOL \citep{BYOL} contrastive learning method over a sample of $\sim$10$^{5}$ radio source RGZ images from the VLA FIRST survey \citep{Becker1995}. The resulting self-supervised model was then fine-tuned to classify FRI/FRII radio galaxies from the VLA FIRST survey, as listed in the \textit{Mirabest} dataset \citep{Mirabest}. The analysis was repeated over a second dataset that include data from the MeerKAT MIGHTEE survey. Both analyses indicated an increase in classification accuracy (ranging from few percent to 8\% for MIGHTEE) over the same model trained in a completely supervised way. \cite{Mohale2023} carried out a similar FRI/FRII classification analysis over the \textit{Mirabest} dataset, using self-supervised models, previously pre-trained over the \textit{ImageNet-1k} (natural images), RGZ (radio galaxy images), Galaxy Zoo DECaLS (optical galaxy images) datasets. Using a KNN classifier evaluation, they found that the model pre-trained on RGZ outperforms the others by a considerable margin (5\% to 10\% improvement in accuracy).
\cite{Hossain2023} carried out the same analysis with both BYOL and SimCLR \citep{SimCLR} self-supervised models but using Group Equivariant Convolutional Neural Network (G-CNN) backbones to make models invariant to different isometries (translation, rotation, mirror reflection). They pre-trained self-supervised models on the RGZ dataset and fine-tuned them on \textit{Mirabest} dataset, obtaining FRI/FRII classification accuracies around 95\%-97\%\footnote{The observed metric differences between BYOL and SimCLR pre-trained models are not significant (below 1\%) given the reported uncertainties.}, improving by $\sim$2\% the fully supervised baseline.\\With respect to previous studies, we focused more on SKA precursor data, building self-supervised models over large samples of unlabelled images, extracted from ASKAP and MeerKAT radio maps in two different modes: (1) "source-centered" mode, e.g. images centred and zoomed over catalogued  sources (as in previous studies); (2) "blind" or "random" mode, e.g. images with arbitrary fixed size extracted from the entire map, without any source position awareness. 
We assessed trained self-supervised models using significantly larger test datasets compared to previous studies, across three representative radio source analysis tasks: radio source morphology classification, radio source instance segmentation, and search for radio objects with peculiar morphology. This study aims to quantify the benefits of self-supervision for the radio domain, providing ready-to-use foundational models that can be exploited in SKA precursor or other radio surveys as feature extractors for similar analysis or to tackle completely new tasks.\\ 
The paper is organized as follows. In Section~\ref{sec:ssl} we describe the contrastive learning model considered, along with the training datasets, data pre-processing and training methodologies adopted. In Sections~\ref{sec:morph_classification}, \ref{sec:source-detection}, 
and \ref{sec:anomaly} we studied how the trained self-supervised models perform in the three selected analysis scenarios, reporting performances achieved against benchmark supervised models. Finally, in Section~\ref{sec:summary} we summarize the obtained results and discuss future steps. 

\begin{figure*}[!htb]
\centering%
\subtable[]{\includegraphics[scale=0.25]{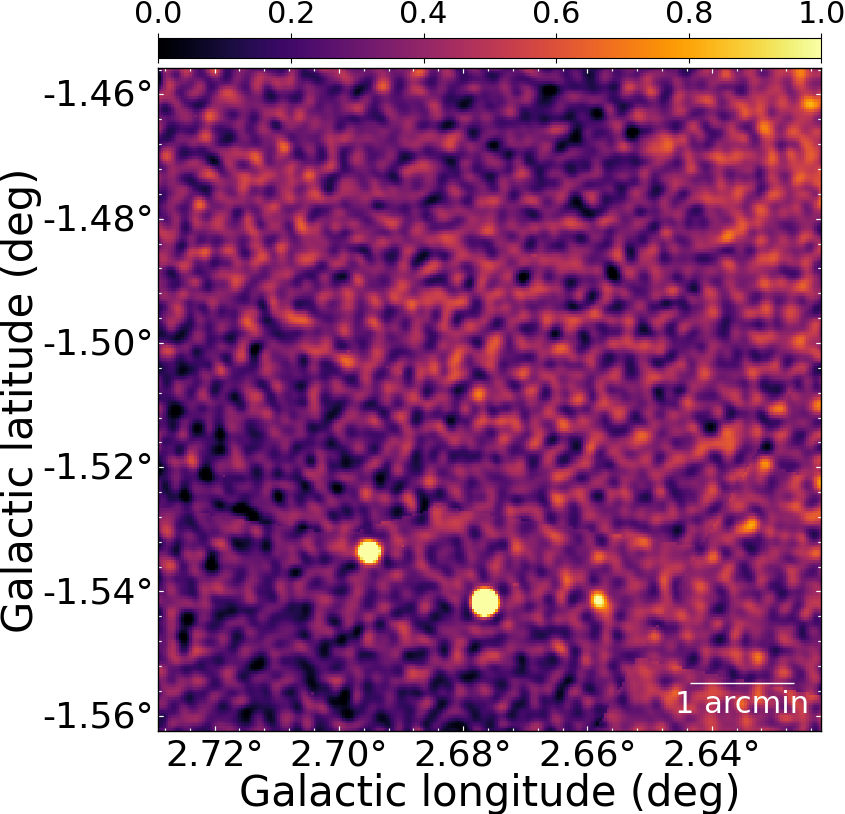}\label{fig:sample_images_1}}
\subtable[]{\includegraphics[scale=0.25]{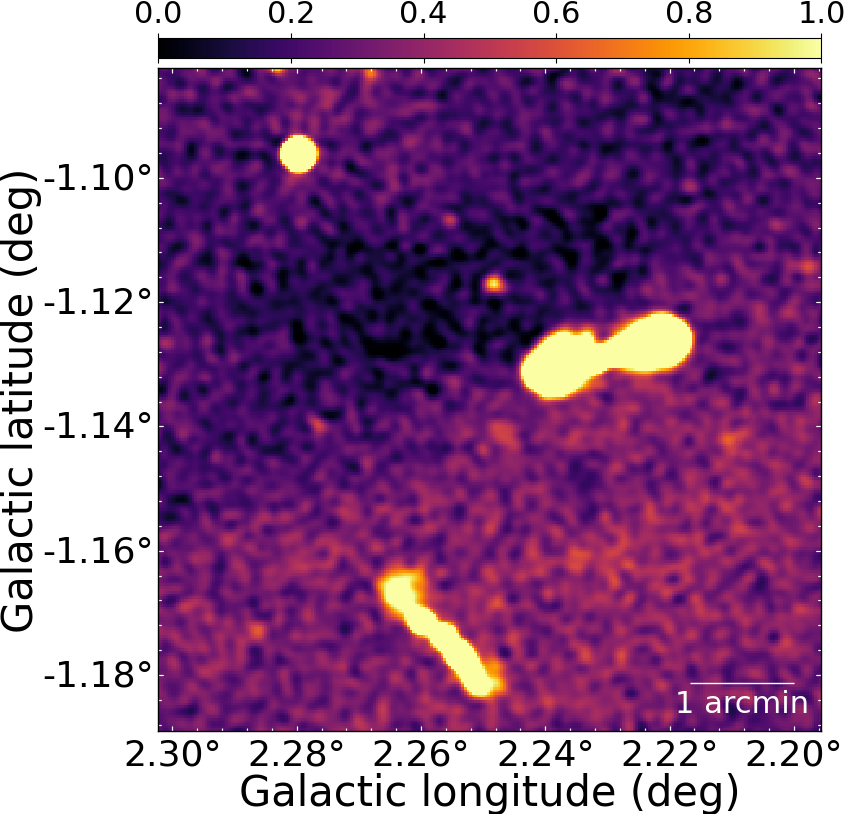}\label{fig:sample_images_2}}%
\subtable[]{\includegraphics[scale=0.25]{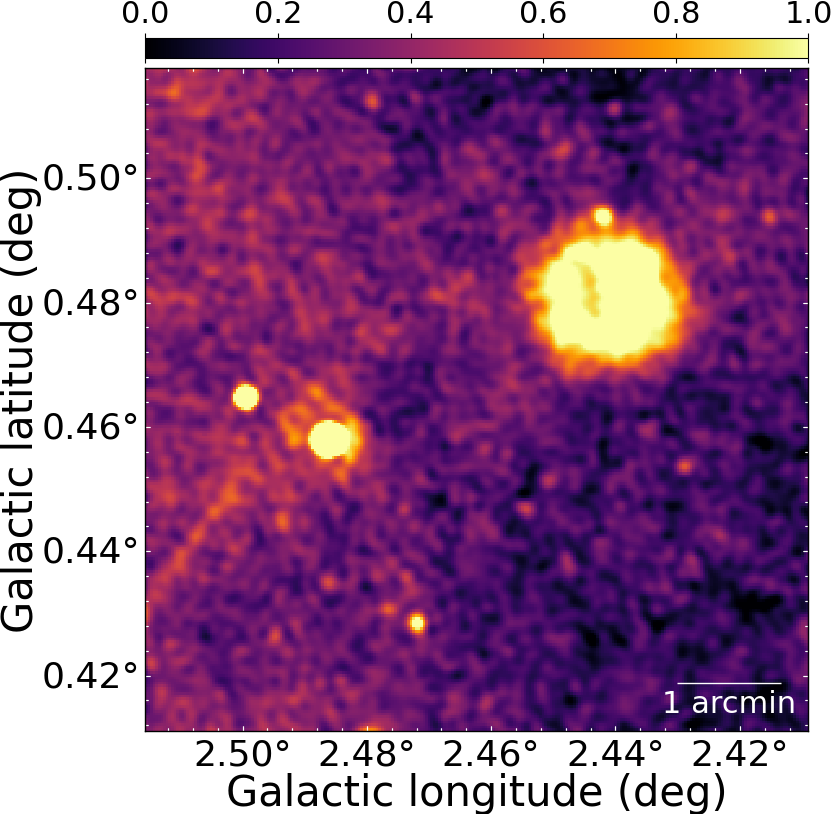}\label{fig:sample_images_3}}%
\\%
\vspace{-0.2cm}
\subtable[]{\includegraphics[scale=0.25]{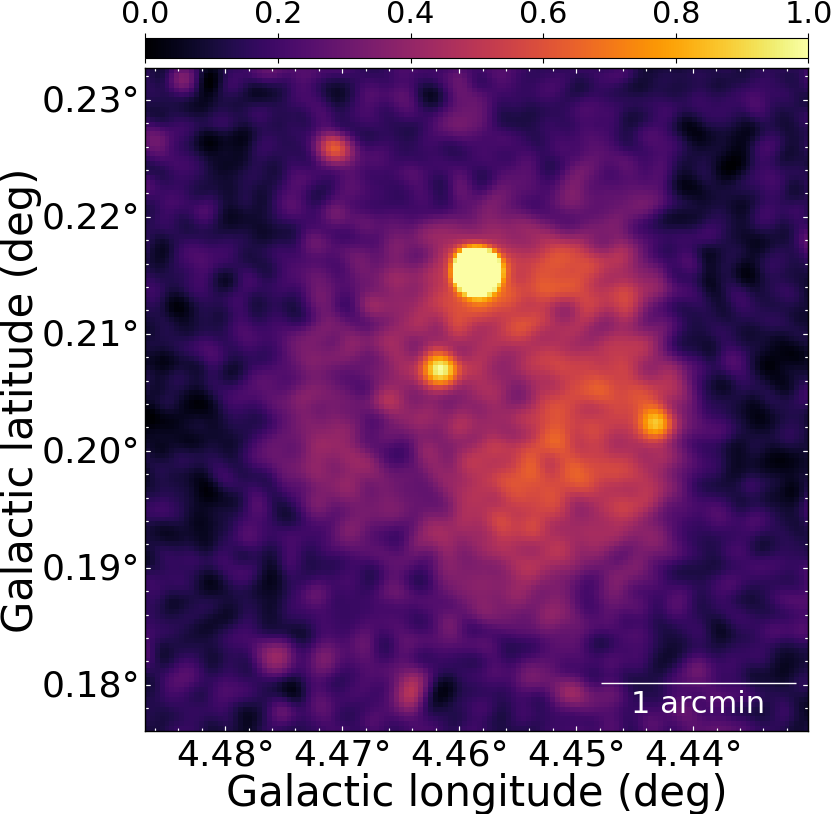}\label{fig:sample_images_4}}%
\subtable[]{\includegraphics[scale=0.25]{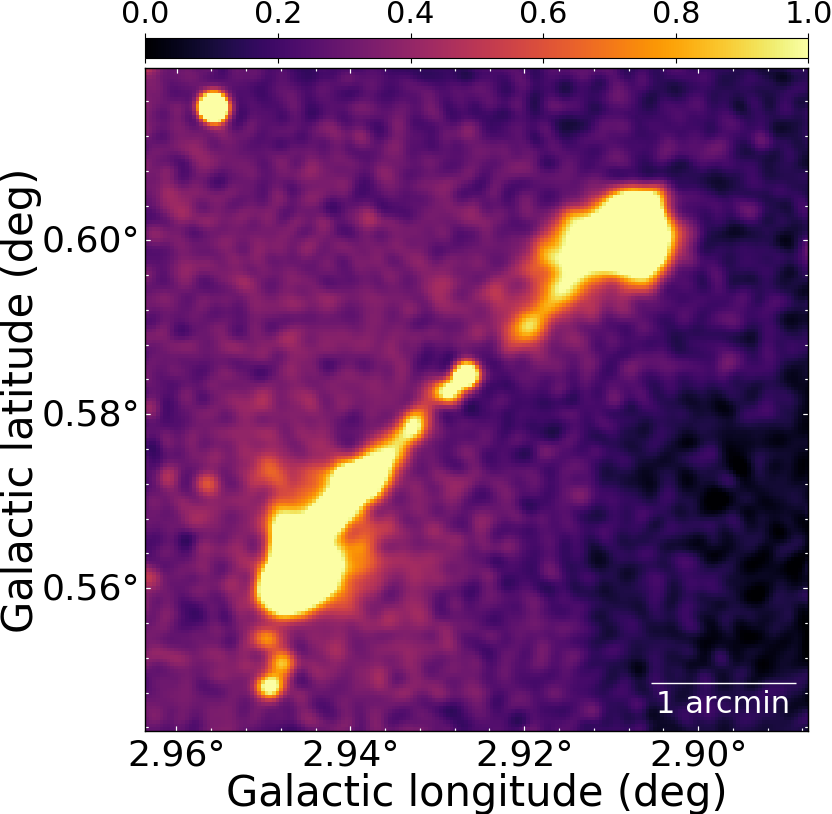}\label{fig:sample_images_5}}%
\subtable[]{\includegraphics[scale=0.25]{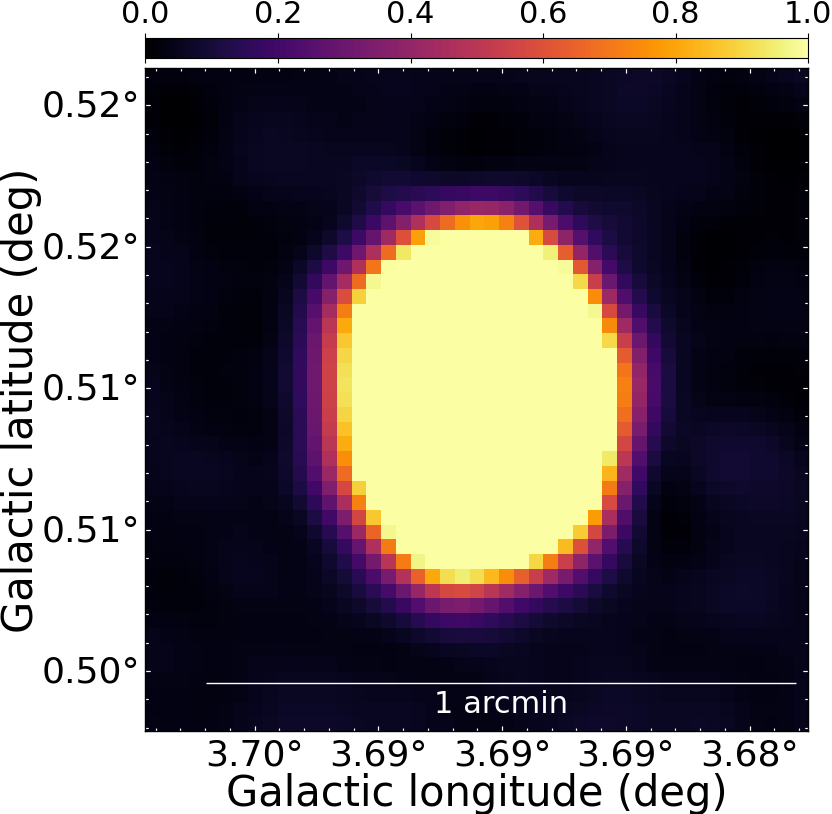}\label{fig:sample_images_6}}%
\\%
\vspace{-0.2cm}
\subtable[]{\includegraphics[scale=0.25]{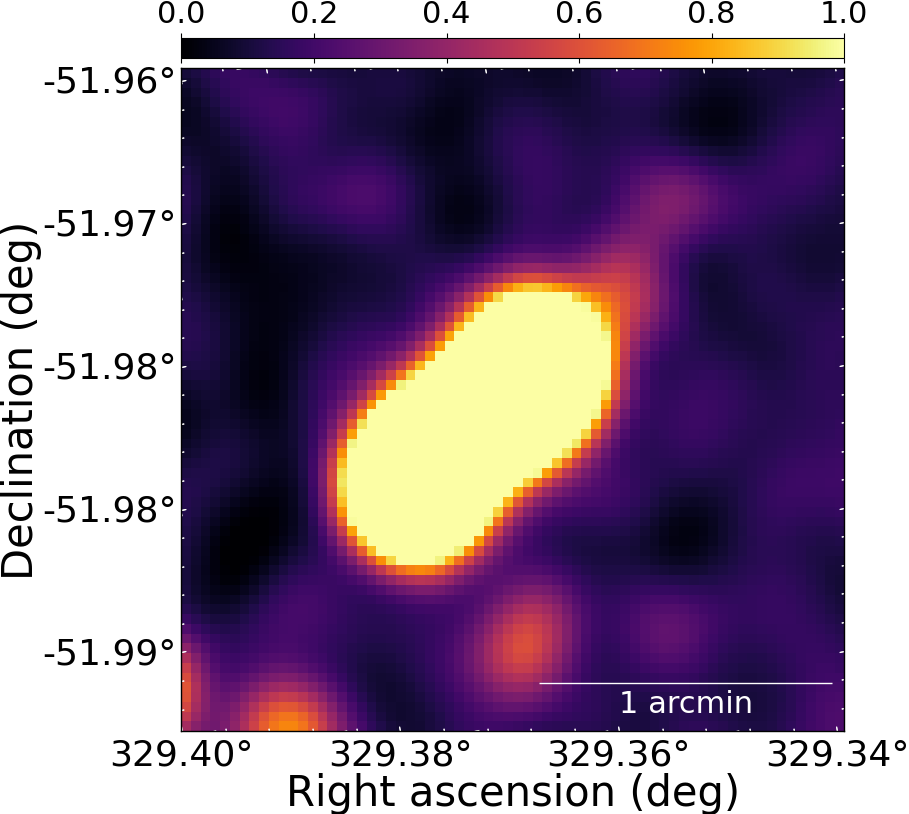}\label{fig:sample_images_7}}%
\subtable[]{\includegraphics[scale=0.25]{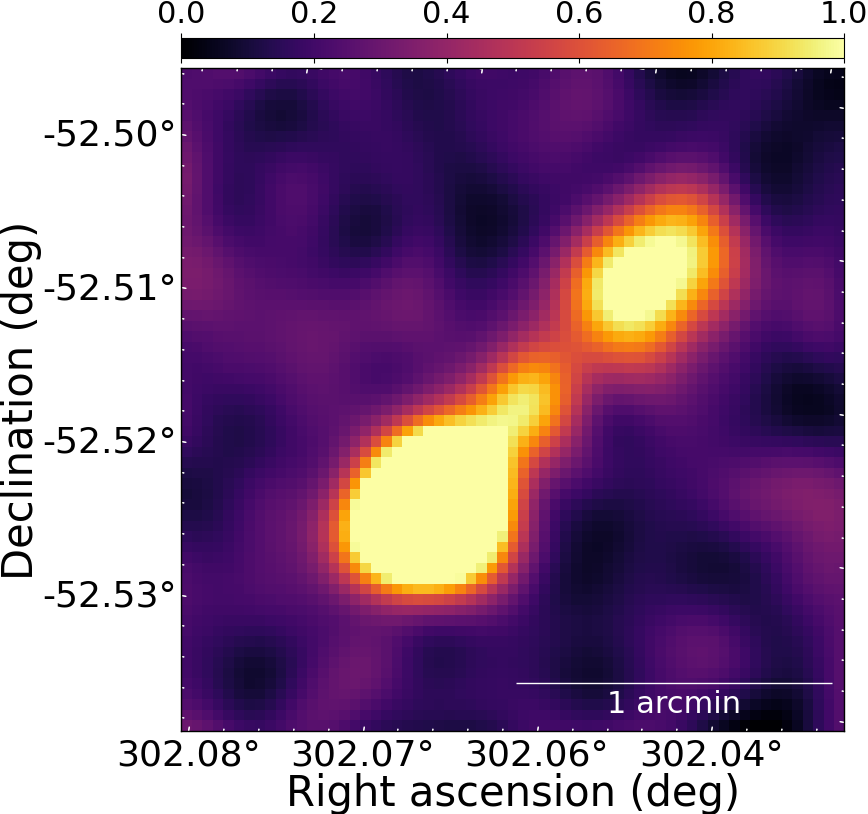}\label{fig:sample_images_8}}%
\subtable[]{\includegraphics[scale=0.25]{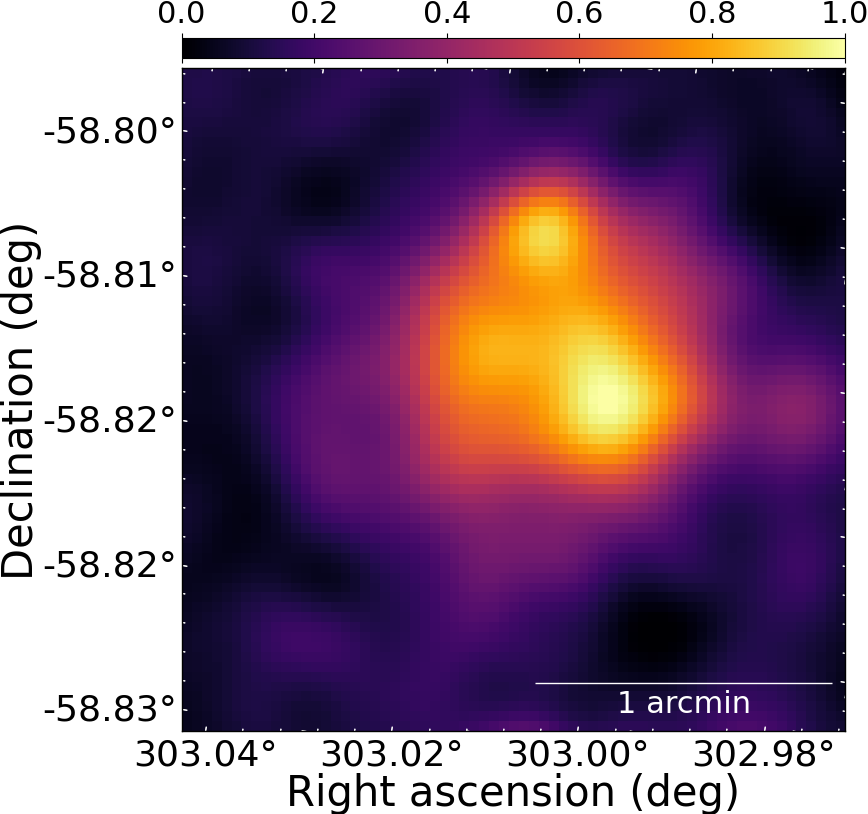}\label{fig:sample_images_9}}%
\vspace{-0.2cm}%
\caption{Representative examples of images from the \texttt{hulk\_smgps} (top panels), \texttt{banner\_smgps} (middle panels) and \texttt{hulk\_emupilot} (bottom panels) datasets. A zscale transform was applied to all images for visualization scopes. Top panels: sample images containing only compact sources (Fig.~\ref{fig:sample_images_1}), or multiple extended sources (Figures \ref{fig:sample_images_2} and \ref{fig:sample_images_3}). Middle panels: sample source with diffuse morphology (Fig.~\ref{fig:sample_images_4}), a multi-component extended source exhibiting typical radio galaxy morphology (Fig.~\ref{fig:sample_images_5}), a single-component extended source with a roundish morphology (Fig.~\ref{fig:sample_images_6}). Bottom panels: sample sources with FR-I (Fig.~\ref{fig:sample_images_7}), FR-II (Fig.~\ref{fig:sample_images_8}) and peculiar (Fig.~\ref{fig:sample_images_9}) classification.}%
\label{fig:sample_images}
\end{figure*}

\section{Self-supervised learning of radio data}
\label{sec:ssl}

\subsection{Contrastive Learning Model}
Fig.~\ref{fig:ssc_schema} illustrates how self-supervised learning can be used for radio data analysis. Initially, a self-supervised framework (indicated by the red block) is trained on large samples of unlabelled image data. Subsequently, the resulting model backbone and data representation (or latent space vector) can be utilized for various downstream tasks, such as data inspection or anomaly detection, typically employing dimensionality reduction methods. Furthermore, the model can be applied to source detection and classification analysis using new labelled datasets. In this study, we used \textit{SimCLR} as the self-supervised framework for our analysis.\\
SimCLR \citep{SimCLR} is a simple yet widely used popular self-supervised learning framework. It learns data representations by maximizing the similarity between augmented views of the same input data (\emph{positive examples}) relative to augmented views of different input data within the same training batch (\emph{negative examples}). The architecture of SimCLR, depicted in Fig.~\ref{fig:ssc_schema}, consists of two main components: a base encoder network $f$, which is typically a \textit{ResNet} network \citep{ResNet}, and a small projection head network $g$, which is typically a Multi-Layer Perceptron (MLP) with one or two layers.
Input images $\mathbf{x}_{k}$ (k=1,\dots,N) in a given batch sample of size $N$ are first processed to produce two augmented views (or positive pair) $\hat{\mathbf{x}}_{2k-1}$ and $\hat{\mathbf{x}}_{2k}$, by randomly applying multiple transformations from a specified transform set $\mathcal{T}$. The encoder network, also denoted as the backbone model throughout the paper, extracts representation vectors $\mathbf{h}_{2k-1}= f(\mathbf{x}_{2k-1})$ and $\mathbf{h}_{2k}= f(\mathbf{x}_{2k})$ from each augmented data pair. The projector network maps the representations to a space where a contrastive loss is applied, obtaining vectors $\mathbf{z}_{2k-1}= g(\mathbf{h}_{2k-1})$ and $\mathbf{z}_{2k}= g(\mathbf{h}_{2k})$. The contrastive loss $\mathcal{L}$, which is minimized during model training, is defined as:
\begin{eqnarray}
\mathcal{L}&=&\frac{1}{2N}\sum_{k=1}^{N}[l_{2k-1,2k} + l_{2k,2k-1}] \\%
l_{i,j}&=&-\log{\frac{\exp(\text{sim}(\mathbf{z}_{i},\mathbf{z}_{j})/\tau)}{\sum_{k=1}^{2N} \mathds{1}_{k\neq i} \exp(\text{sim}(\mathbf{z}_{i},\mathbf{z}_{k})/\tau)}}
\end{eqnarray}
where $l_{i,j}$ is the normalized temperature-scaled cross entropy loss (NT-Xent), $\mathds{1}_{k\neq i}$=1 if $k$=$i$ (equal to 0 otherwise), $\tau$ is a temperature parameter, and $\text{sim}(\mathbf{z}_i,\mathbf{z}_j)$ is the pair-wise similarity between vectors $\mathbf{z}_i$ and $\mathbf{z}_j$, defined as: 
\begin{equation}
\text{sim}(\mathbf{z}_i,\mathbf{z}_j)= \frac{\mathbf{z}_{i}^{T}\mathbf{z}_j}{\parallel\mathbf{z}_i\parallel\;\parallel\mathbf{z}_j\parallel }
\end{equation}

\begin{table}[htb]
\centering%
\scriptsize%
\caption{Summary information of datasets used for SimCLR model training. The number of images $n_{img}$ is reported in column (2). The image size $s_{img}$ is reported in column (3). $s_{img}$ is fixed for all images in the \texttt{hulk\_smgp} and \texttt{hulk\_emupilot} datasets, while $s_{img}$ is not fixed and depends on the source size $s_{\text{source}}$ (equivalent to the maximum source bounding box dimension) in the \texttt{banner\_smgps} and \texttt{banner\_emupilot} datasets. For these datasets, we report the average, minimum and maximum source sizes in columns (4), (5) and (6), respectively. Images from all datasets are eventually resized to a fixed size for model training and testing (see Section~\ref{subsec:simclr-preproc}).}
\begin{tabular}{llcccc}
\hline%
\hline%
&  & & \multicolumn{3}{c}{s$_{\text{source}}$} \\%
\cline{4-6}%
 \emph{Dataset} & n$_{img}$ &  s$_{img}$ & $\langle s_{\text{source}}\rangle$ & $s_{\text{source}}^{min}$  & $s_{\text{source}}^{max}$ \\%
 & & \scriptsize{(pix)} & \scriptsize{(arcmin)} & \scriptsize{(arcsec)} & \scriptsize{(arcmin)}\\%
\hline%
\texttt{hulk\_smgps} & 178,057 &  & $-$ & $-$ & $-$\\%
\texttt{hulk\_emupilot} & 55,773 & \multirow{-2}{*}{256$\times$256} & $-$ & $-$ & $-$\\%
\hline%
\texttt{banner\_smgps} & 17,062 &  & 1.3 & 11.3 & 24.7\\%
\texttt{banner\_emupilot} & 10,414 & \multirow{-2}{*}{1.5$\times$s$_{\text{source}}$} & 1.2 & 18.8 & 7.8\\%
\hline%
\hline%
\end{tabular}
\label{tab:simclr_datasets}
\end{table}

\subsection{Datasets}
\label{sec:ssc-dataset}
We created the following unlabelled datasets for training SimCLR:
\begin{enumerate}
\item Two distinct datasets were generated using data from the SARAO MeerKAT Galactic Plane Survey (SMGPS) \citep{Goedhart2024}, which covers a large portion of the 1st, 3rd and 4th Galactic quadrants (l=2$^{\circ}$-61$^{\circ}$, 251$^{\circ}$-358$^{\circ}$, $|b|<$1.5$^{\circ}$) in the L-band (886-1678 MHz). The survey has an angular resolution of 8" and a noise rms of $\sim$10-20 $\mu$Jy/beam at 1.3 GHz:
\begin{itemize}
\item \texttt{hulk\_smgps}: A collection of 178,057 radio images, each of fixed size (256$\times$256 pixels, equivalent to $\sim$6.4'$\times$6.4'), extracted from SMGPS 1.28 GHz integrated intensity maps. This dataset was created by assuming a sliding window that traverses the entire surveyed area with a shift size equal to half the frame size, resulting in a 50\% overlap among frames. 
The image size was chosen to be large enough to encompass the most extended radio galaxies that might be located in the cutout\footnote{Out of $\sim$5800 catalogued sources that were labelled as candidate radio galaxies on the basis of their radio morphology, only one was found to have a size (7.4') larger than the chosen image cutout (6.4').};
\item \texttt{banner\_smgps}: A collection of 17,062 radio images extracted from SMGPS 1.3 GHz integrated maps, each centered around sources listed in the SMGPS extended source catalogue \citep{Bordiu2024}. The size of the images varies across the dataset and is set to 1.5 times the size of the source bounding box. The radio sources in this dataset exhibit different morphologies, including single-island, multi-island, and diffuse sources.
\end{itemize}
\item Two distinct datasets were generated using data from the ASKAP EMU pilot survey \citep{Norris2021}, which covered approximately 270 deg$^{2}$ of the Dark Energy Survey area, achieving an angular resolution of 11" to 18" and a noise rms of $\sim$30 $\mu$Jy/beam at 944 MHz:
\begin{itemize}
\item \texttt{hulk\_emupilot}: A collection of 55,773 radio images, each of fixed size (256$\times$256 pixels, equivalent to $\sim$8.5'$\times$8.5'), extracted from ASKAP EMU pilot 944 MHz integrated map.
The images were extracted using a sliding frame that traversed the entire mosaic with a shift size equal to half the frame size, resulting in a 50\% overlap among frames. 
\item \texttt{banner\_emupilot}: A collection of 10,414 radio images extracted from ASKAP EMU pilot 944 MHz integrated map, each centered around extended sources listed in the pilot source catalogue compiled by \cite{Gupta2024}. The size of the images varies across the dataset and is set to 1.5 times the size of the source bounding box. The radio sources in this dataset exhibit different morphologies, including FR-I ($\sim$6\%), FR-II ($\sim$54\%), FR-x ($\sim$14\%), single-peak resolved ($\sim$23\%) radio galaxies. $\sim$3\% of the sources present a rare morphology not fitting into the previously mentioned categories. 
\end{itemize}
\end{enumerate}

\begin{table}[htb]
\centering%
\footnotesize%
\caption{List of augmentations used in SimCLR model training. In column (2) we reported the transform parameter values. In column (3) we reported the probability used to apply the transform in the augmentation pipeline, e.g. 1.0 means the transform is always applied to all input images.}
\begin{tabular}{lll}
\hline%
\hline%
Augmentation & Parameters & Probability\\%
\hline%
\emph{RandomCropResize} & \texttt{crop\_size}=[0.8,1.0] & 1.0\\%
\emph{ColorJitter} & \texttt{brightness}=0.8 & 0.8\\%
& \texttt{contrast}=0.8 & \\%
& \texttt{saturation}=0.8 & \\%
& \texttt{hue}=0.2 & \\%
& \texttt{strength}=0.5 & \\%
\emph{HorizontalFlip} & $-$ & 0.33\\%
\emph{VerticalFlip} & $-$ & 0.33\\%
\emph{Rotate} & \texttt{angle}=\{90,180,270\} & 0.5\\%
\emph{Blur} & \texttt{sigma}=[1,3] & 0.1\\%
\emph{RandomThresholding} & \texttt{percentileThr}=[40,60] & 0.5\\%
\hline%
\hline%
\end{tabular}
\label{tab:simclr_augmentations}
\end{table}

Datasets extracted in a blind mode (e.g. without any previous knowledge of the source location and morphology) can be constructed rapidly, potentially reaching substantial sizes (up to millions of images) when using future full-sky surveys. Without additional selection processes, these datasets tend to be largely unbalanced, predominantly comprising frames composed entirely of compact sources. The \texttt{hulk\_smgps} dataset also comprises frames with large-scale diffuse emission, including background or portions of very extended sources located along the Galactic plane. For simplicity, we have labelled them as \texttt{hulk}. In contrast, "smarter" datasets centered on selected source positions typically have smaller sizes, requiring significant efforts (catalogue production and source type annotation) for construction. We have labelled them as \texttt{banner}.
Indeed, one goal of this work is evaluating differences and benefits of both kind of datasets over different analysis tasks. Summary information for all produced datasets is reported in Table~\ref{tab:simclr_datasets}. In Fig.~\ref{fig:sample_images} we display sample images from the \texttt{hulk\_smgps} (top panels), \texttt{banner\_smgps} (middle panels) and \texttt{banner\_emupilot} (bottom panels) datasets.


\begin{table}[htb]
\centering%
\footnotesize%
\caption{List of hyperparameters used in SimCLR model training.}
\begin{tabular}{lll}
\hline%
\hline%
Hyperparameter & Value\\%
\hline%
Encoder model & ResNet18\\
Projector model size & 2 layers (256-128)\\
Optimizer & Adam\\%
Batch size & 128\\%
Learning Scheduler & Linear warmup + Cosine decay\\%
Warmup epochs & 10\\%
Learning rate (warmup target) & 0.1\\%
Epochs & 100 (\texttt{hulk\_smgps} datasets)\\%
& 500 (\texttt{banner\_smgps} dataset)\\%
& 100 (\texttt{hulk\_emupilot} dataset)\\%
& 500 (\texttt{banner\_emupilot} dataset)\\%
\hline%
\hline%
\end{tabular}
\label{tab:simclr_hyperparameters}
\end{table}

\begin{figure*}[htb]
\centering%
\subtable[\texttt{1C-1P}]{\includegraphics[scale=0.25]{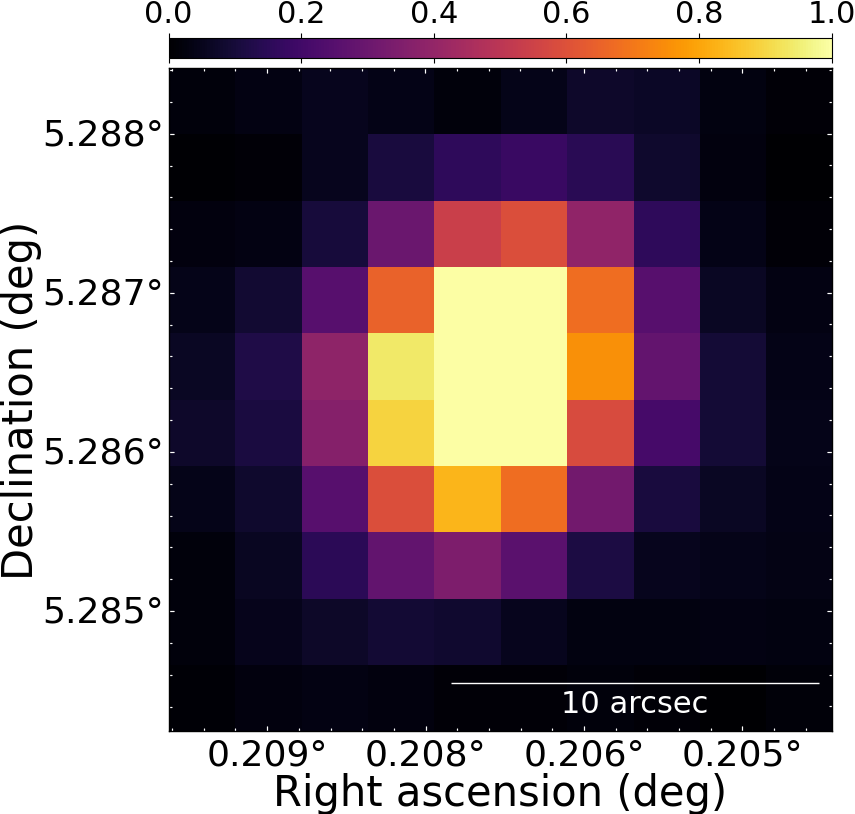}}%
\subtable[\texttt{1C-2P}]{\includegraphics[scale=0.25]{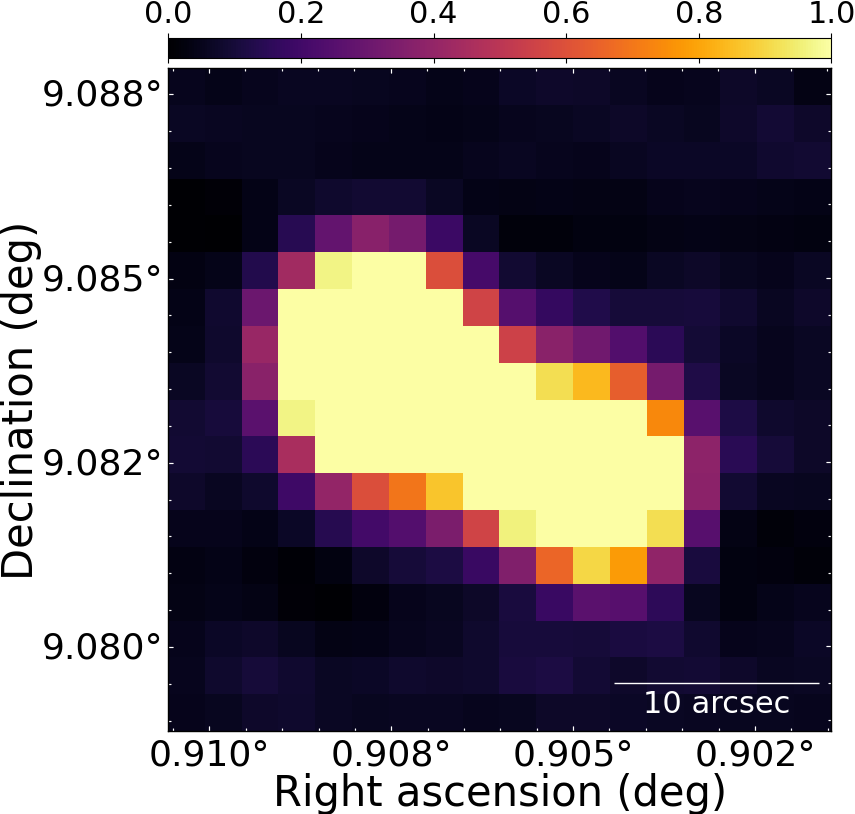}}%
\subtable[\texttt{1C-3P}]{\includegraphics[scale=0.25]{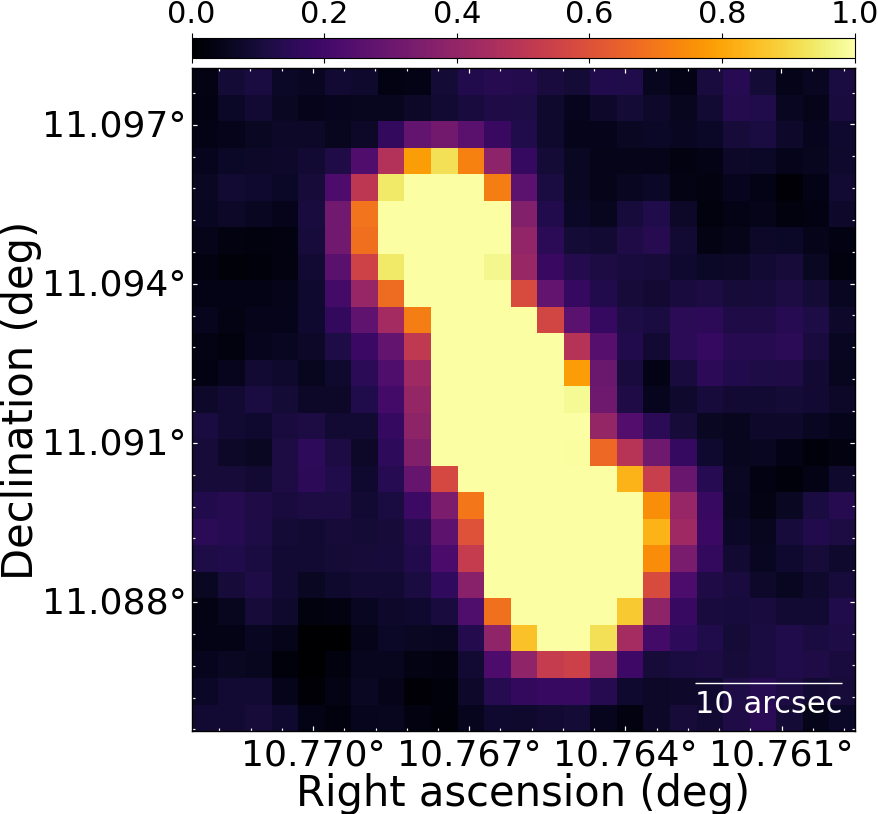}}\\%
\vspace{-0.2cm}
\subtable[\texttt{2C-2P}]{\includegraphics[scale=0.25]{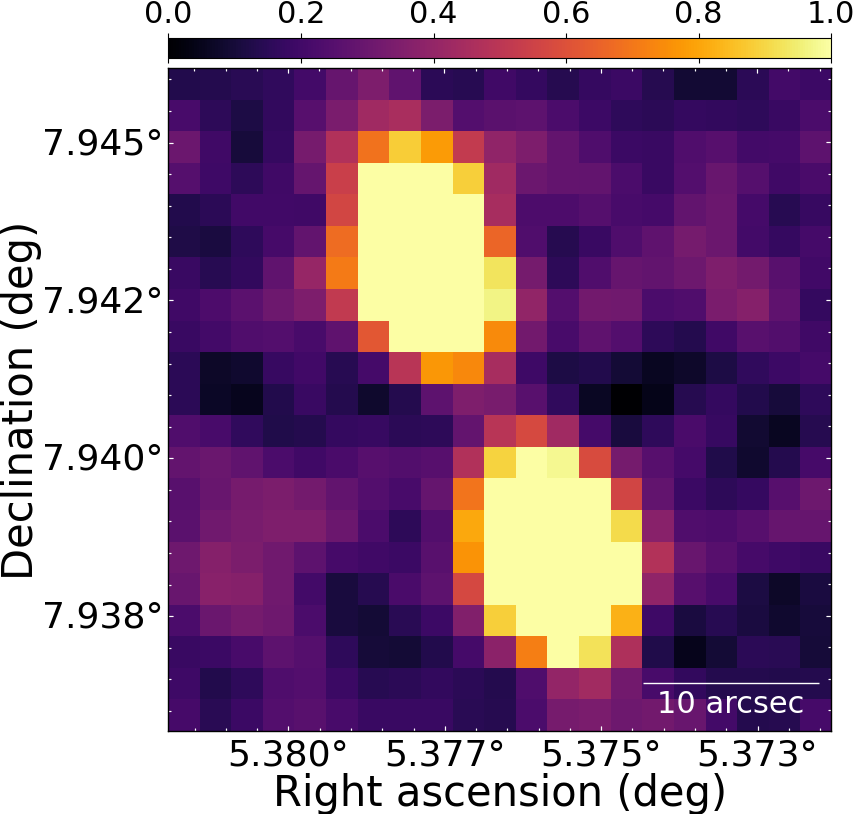}}%
\subtable[\texttt{2C-3P}]{\includegraphics[scale=0.25]{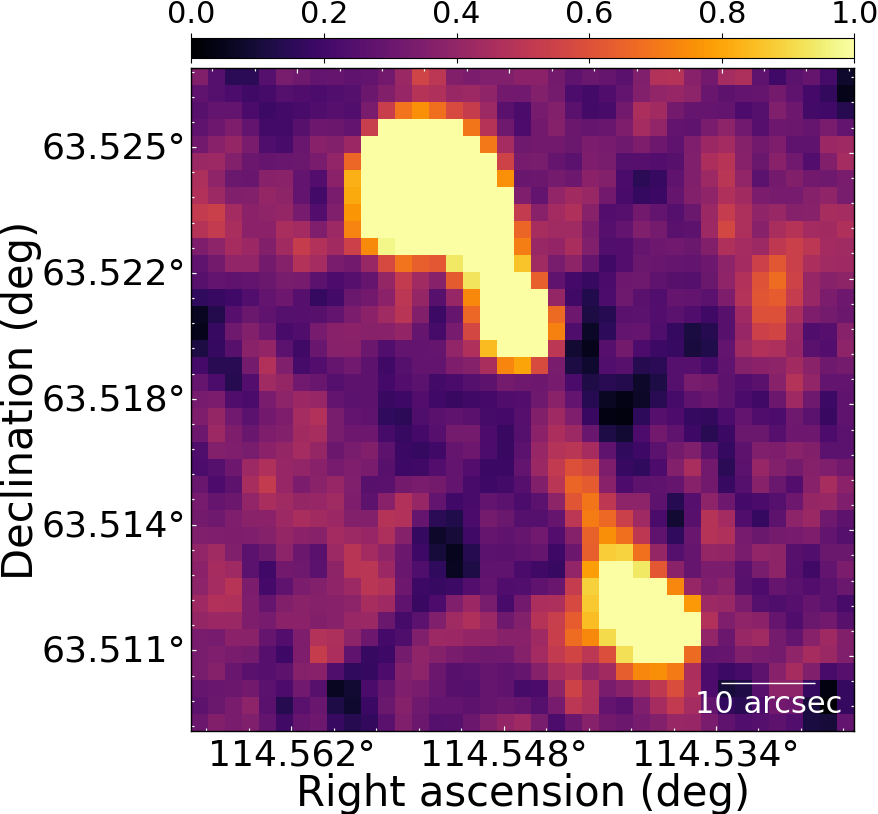}}%
\subtable[\texttt{3C-3P}]{\includegraphics[scale=0.25]{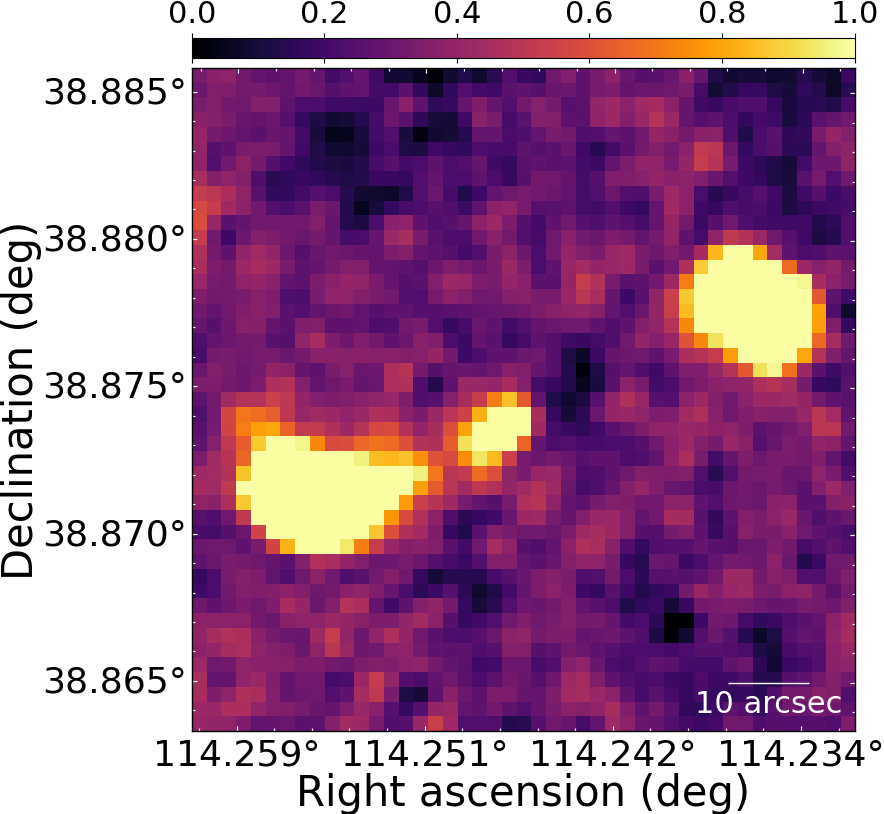}}%
\caption{Sample images from the RGZ dataset with representative sources of different morphological classes (reported below each frame). A zscale transform was applied to all images for visualization scopes.}%
\label{fig:rgz_source_examples}
\end{figure*}

\subsection{Data pre-processing and augmentation}
\label{subsec:simclr-preproc}
For the training and inference stages, we applied these pre-processing steps to input images:
\begin{itemize}
\item Grayscale images were converted to 3-channels. Each channel was processed differently from others, applying the following transformations:
\begin{itemize}
\item \emph{Channel 1}: sigma-clipping ($\sigma_{low}$=5, $\sigma_{up}$=30);
\item \emph{Channel 2}: zscale transform (contrast=0.25);
\item \emph{Channel 3}: zscale transform (contrast=0.4).
\end{itemize}
\item Each channel was independently normalized to a [0,1] range using a \emph{MinMax} transformation;
\item Finally, each channel was resized to a 224$\times$224 size in pixels.
\end{itemize}
A key aspect when training contrastive learning models is the choice of applied data augmentation steps to make the model invariant with respect to non-physical properties or to features not related to the radio sources. We applied the following augmenters to the data sequentially:
\begin{itemize}
\item \emph{RandomCropResize}: randomly crop input images to size \texttt{crop\_size} $\times$ image size, and resize data to the original size. \texttt{crop\_size} was randomly varied in the range [0.8, 1.0];
\item \emph{ColorJitter}: apply a colour jitter transformation using all three image channels;
\item \emph{Flip}: random flip images either vertically or horizontally;
\item \emph{Rotate}: rotate images by either 90, 180 or 270 degrees;
\item \emph{Blur}: apply Gaussian blurring to images using a $\sigma$ parameter randomly varied in the range [1,3];
\item \emph{RandomThresholding}: threshold each channel separately using a per-channel percentile threshold randomly varied in the range [40,60].
\end{itemize}
The \emph{RandomThresholding} augmenter was introduced to make the model less dependent on image background features. This stage was not applied when training over the \texttt{banner} datasets, as images in this dataset are already zoomed on radio sources, and thus the background would likely not be estimated correctly. Furthermore, not all augmenters are applied to every image in the training dataset. In Table~\ref{tab:simclr_augmentations} we provide a summary of augmenter steps used in the pre-processing pipeline with their parameters, including the probability with which each data transform is applied to images. With respect to \cite{SimCLR}, we reduced the fraction of random cropping allowed to avoid cutting out relevant details of extended sources from the resulting image.

\subsection{Model training}
We trained a SimCLR model on each of the four datasets described in Section~\ref{sec:ssc-dataset}, using the hyperparameters listed in Table~\ref{tab:simclr_hyperparameters}. We will refer to them using their training dataset name: \texttt{hulk\_smgps}, \texttt{banner\_smgps}, \texttt{hulk\_emupilot}, and \texttt{banner\_emupilot}.
A fourth model, referred to as \texttt{smart\_hulk\_smgps} hereafter, was trained in two steps, first on the \texttt{hulk\_smgps} dataset and then on the \texttt{banner\_smgps} dataset. The final model weights from the first step were used as initialization for the second step. For all models, we used a \emph{ResNet18} \citep{ResNet} encoder and a 2-layer projector with 256 and 128 neurons, respectively.\\Following \cite{SimCLR}, all training runs began with a linear warm-up phase lasting 10 epochs, after which we switched to a cosine learning rate decay strategy. In total, we trained models for 500 epochs on the \texttt{banner\_smgps} and \texttt{banner\_emupilot} datasets. A smaller total number of epochs (100) was used when training over the larger \texttt{hulk\_smgps} and \texttt{hulk\_emupilot} datasets to reduce computational costs.\\Training runs were performed on three different computing server nodes, each equipped with a GPU device:
\begin{itemize}
\item Node A: 48 cores (Intel Xeon Gold 6248R CPU, 3.00 GHz), 512 GB RAM, NVIDIA Quadro RTX 6000 (24 GB)
\item Node B: 24 cores (Intel Xeon Silver 4410Y, 2.00 GHz), 256 GB RAM, NVIDIA A30 (24 GB) 
\item Node C: 36 cores (Intel Xeon CPU E5-2697 v4, 2.30 GHz), 128 GB RAM, NVIDIA Tesla V100 (16 GB)
\end{itemize}
Typical training times over the \texttt{hulk\_smgps} dataset are of the order of $\sim$6.7 hours/epoch on nodes A/B, and $\sim$12.5 hours/epoch on node C.

\begin{figure}[htb]
\centering%
\includegraphics[scale=0.42]{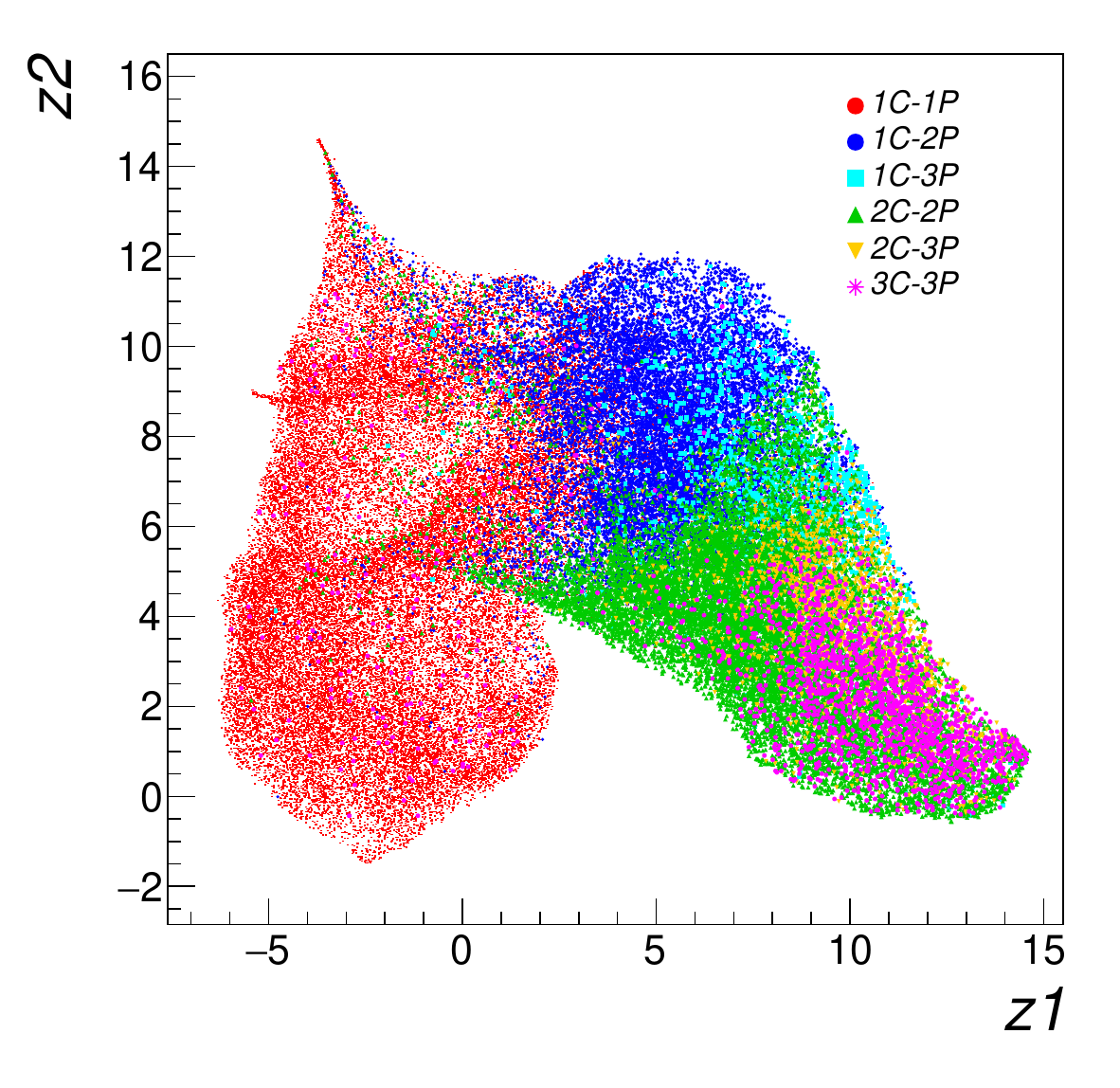}%
\caption{2D UMAP projection of the data representation vector (size=512) produced by the trained \texttt{smart\_hulk\_smgps} model on the RGZ dataset.}%
\label{fig:rgz_umap}
\end{figure}

\section{Task I: Classification of radio source morphology}
\label{sec:morph_classification}
In this section, we quantitatively evaluate the learned self-supervised representation on a source morphology classification problem.\\Morphological classification plays a pivotal role in radio astronomy, enabling scientists to gain insights into the underlying source nature from the observed shape and structures. The majority of existing works in the radio image domain are targeted for extragalactic science objectives, focusing on classification of radio galaxies (see for example \citealt{Aniyan2017}, \citealt{Ma2019}, or \citealt{Ndungu2023} for a recent review) in different morphological classes: \texttt{compact}, \texttt{FR-I}, \texttt{FR-II}, \texttt{bent-tailed} (including WAT\footnote{Wide-angle tail} and NAT\footnote{narrow-angle tail} population), \texttt{irregular} (including, for example, X-shaped or ring-like radio galaxies).\\Morphological classification is also an important post-detection stage to filter extracted sources by general morphology (e.g. compact vs extended sources) for specialized source property measurements or other advanced classification analysis. In this context, the adopted source labelling scheme is rather general-purpose and domain-agnostic, suited to be eventually refined afterwards. For example, typical used labels are \texttt{POINT-LIKE}, \texttt{RESOLVED}, \texttt{COMPACT}, \texttt{EXTENDED} or labels that contain information about the number of radio components present in the extracted source (as in \citealt{Wu2019}).\\The analysis carried out in this section falls into the second use-case scenario. This choice is mostly driven by existing datasets. Available annotated datasets for radio galaxy classification (mostly based on VLA FIRST survey data) are, in fact, rather limited in size (e.g. typically <100-200 images per class, <2000 images overall) and would currently prevent us from obtaining a robust evaluation of our self-supervised models over multiple test set realizations. For example, the \textit{Mirabest} dataset \citep{Mirabest} contains 1256 source images of balanced FR-I/FR-II radio galaxy classes, out of which 833 images constitute the "Confident" sample, and the rest (423 images) the "Uncertain" sample. On this dataset, \cite{Slijepcevic2024} reported an improvement of $\sim$3-4\% in classification performance of a self-supervised pre-trained model with respect to a fully supervised model trained from scratch on the "Confident" sample (or on a portion of it). Classification metrics were, however, estimated on the "Uncertain" sample, and hence the observed enhancement is due to less than 20 sources. We, therefore, opted for this work to use a larger dataset (roughly by one order of magnitude) and perform a similar analysis once a larger dataset is assembled within the ASKAP EMU survey.

\subsection{Dataset}
For this analysis, we considered data from the Radio Galaxy Zoo (RGZ) project\footnote{The RGZ project is a crowdsourced science project where both scientists and citizens can classify radio galaxies and their host galaxies from radio and infrared (WISE survey, \citealt{Wright2010}) images presented to users in a web interface} \citep{RGZ}. This includes radio images of size 3'$\times$ 3' from the VLA Faint Images of the Radio Sky at Twenty cm (FIRST) survey (1.4 GHz, angular resolution $\sim$5") \citep{Becker1995}. Radio sources found in these images were labelled into multiple morphological classes, on the basis of the observed number of components (C) and peaks (P) (see \citealt{Wu2019} for more details on the classification schema). Angular size is also available for each source.\\In this analysis, we extracted 82,084 image cutouts around radio sources that have been classified in the RGZ Data Release 1 (DR1) with a consensus level $\ge$0.6 in the following classes: 1C-1P (55.0\%), 1C-2P (20.9\%), 1C-3P (1.9\%), 2C-2P (17.6\%), 2C-3P (2.0\%), 3C-3P (2.5\%). We assumed a cutout size equal to 1.5 $\times$ the source angular size, as listed in the RGZ catalogue. A representative image of each source morphological category is shown in Fig.~\ref{fig:rgz_source_examples}.\\As the full dataset is largely unbalanced towards sources of class morphology 1C-1P, we randomly created $N$=5 balanced train and test sets having 1000 and 600 images per each class, respectively. Both train and test set images were pre-processed as described in Section~\ref{subsec:simclr-preproc} for the SimCLR model training.

\begin{figure}[htb]
\centering%
\includegraphics[scale=0.42]{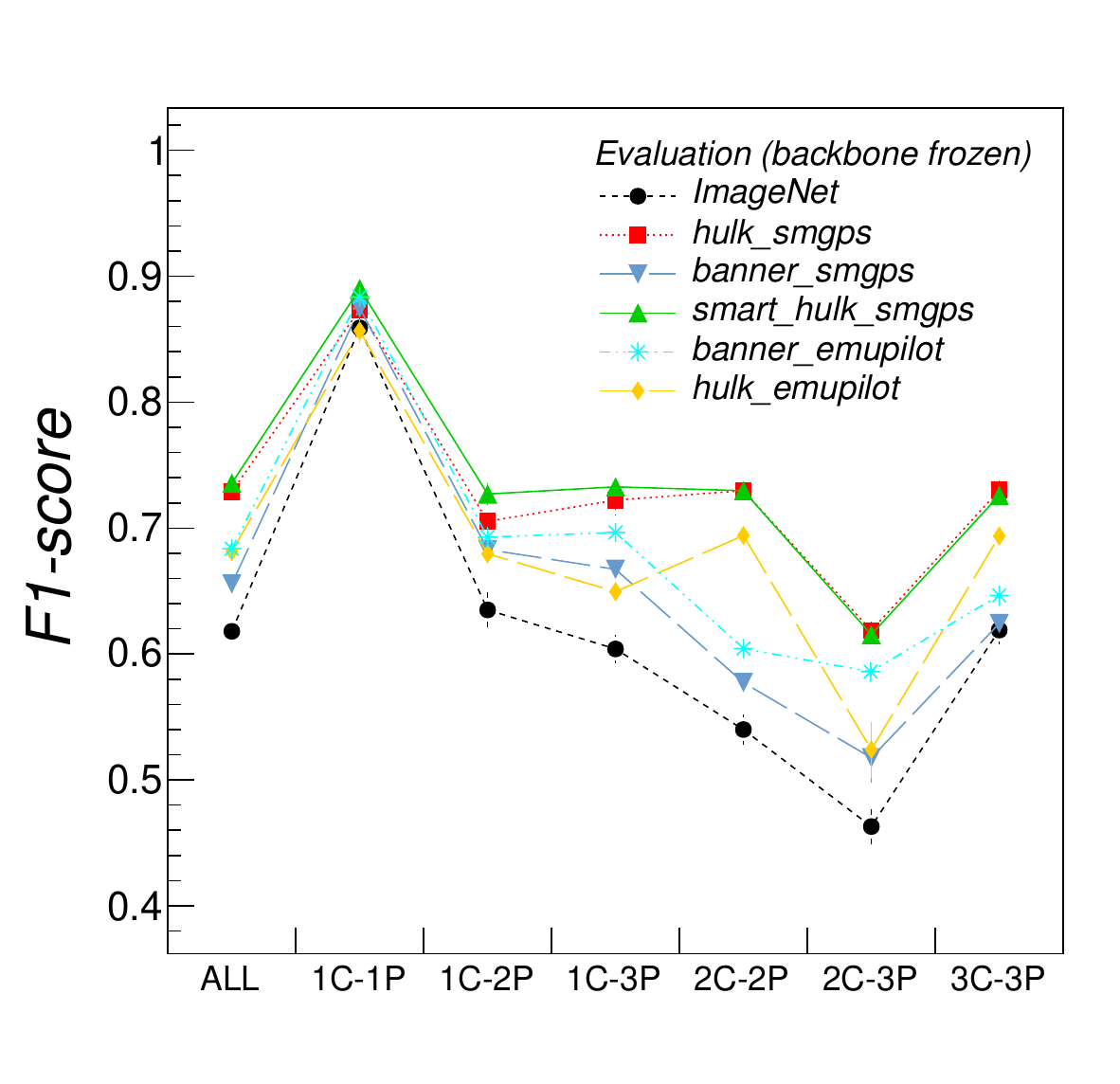}%
\caption{Classification F1-scores obtained for different classes and for all classes cumulatively over RGZ test sets with different pre-trained and frozen backbones: \texttt{hulk\_smgps} (red squares), \texttt{banner\_smgps} (blue inverted triangles), \texttt{smart\_hulk\_smgps} (green triangles), \texttt{hulk\_emupilot} (orange diamonds), \texttt{banner\_emupilot} (cyan asterisks), \texttt{ImageNet} (black dots). The reported values and errors are the F1-score mean and mean error computed over five test sets.}%
\label{fig:rgz_class_eval}
\end{figure}

\begin{figure}[htb]
\centering%
\includegraphics[scale=0.45]{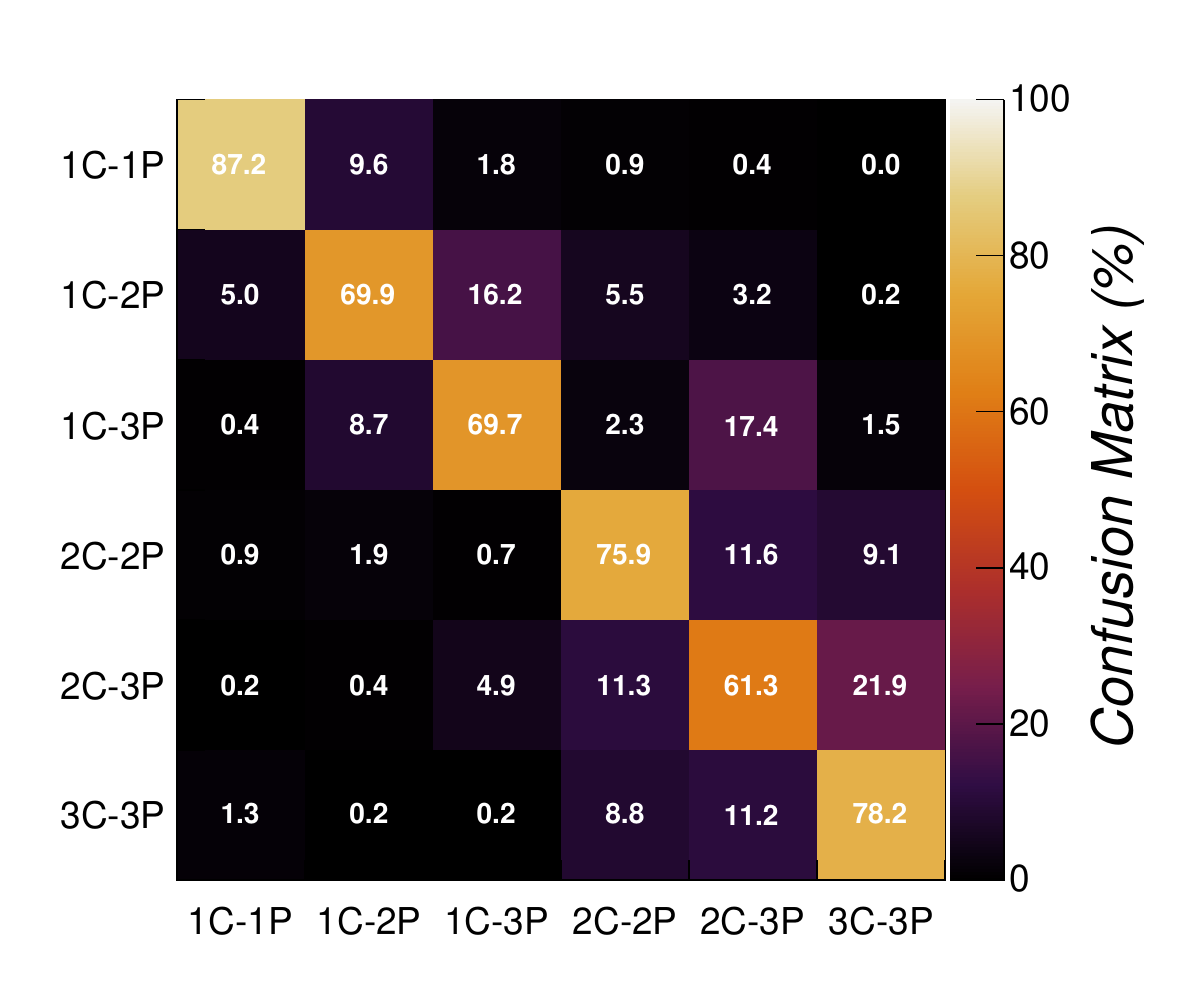}
\caption{Confusion matrix of the source morphology classifier (trained with \texttt{smart\_hulk\_smgps} pre-trained and frozen backbone) obtained over the RGZ test set.}%
\label{fig:rgz_confusion_matrix}
\end{figure}

\subsection{Evaluation of self-supervised representation}
\label{subsec:rgz-eval}
In Fig.~\ref{fig:rgz_umap} we present a two-dimensional projection, obtained with the \emph{Uniform Manifold Approximation and Projection} (UMAP) algorithm, of the representation vector (original size equal to 512) produced by the trained \texttt{smart\_hulk\_smgps} model on the RGZ dataset. As can be observed, the self-supervised model effectively groups sources of different morphological class in distinct areas of the latent space. No isolated clusters are discernible in the projected two-dimensional UMAP feature space, as well as in a PCA scatter plot of top-2 features (not shown here). Nevertheless, these or similar diagnostic plots, can be useful for potentially identifying possible image mislabeling in the dataset, e.g. sources that fall within a region that is predominantly populated by other classes.\\
We carried out a classification analysis using a CNN classifier with a standard architecture: a \emph{ResNet18} backbone (as in the SimCLR model) followed by a classification head. The latter consists of a single layer followed by a softmax activation, representing the predicted probability distribution over the set of classes. To evaluate the quality of the self-supervised representation, we froze the model backbone, setting and fixing its weights to those obtained in the trained SimCLR models, and trained only the classification head on RGZ training datasets for a limited number of epochs (30). 
We considered only rotation and flipping transformations as augmentation steps during the training.
To estimate the achieved classification performances, we used these widely adopted metrics in multi-class classification problems:
\begin{itemize}
\item \emph{Recall} ($\mathcal{R}$): Fraction of sources of a given class that were correctly classified by the model out of all sources labelled in that class, computed as:
\begin{equation*}
\mathcal{R}=\frac{TP}{TP + FN}
\end{equation*}
\item \emph{Precision} ($\mathcal{P}$): Fraction of sources correctly classified as belonging to a specific class, out of all sources the model predicted to belong to that class, computed as:
\begin{equation*}
\mathcal{P}=\frac{TP}{TP+FP}
\end{equation*}
\item \emph{F1-score}: the harmonic mean of precision and recall:
\begin{equation}
\text{F1-score}=2\times\frac{\mathcal{P}\times\mathcal{R}}{\mathcal{P}+\mathcal{R}}
\end{equation}
\end{itemize}
In Fig.~\ref{fig:rgz_class_eval} we report the classification F1-scores obtained on the test set by different self-supervised pre-trained models: \texttt{hulk\_smgps} (red squares), \texttt{banner\_smgps} (blue inverted triangles), \texttt{smart\_hulk\_smgps} (green triangles), \texttt{hulk\_emupilot} (orange diamonds), \texttt{banner\_emupilot} (cyan asterisks). The reported values and their errors are respectively the F1-score mean and mean error computed over the available test sets. These metrics were compared against those obtained with a baseline model pre-trained on the \texttt{ImageNet-1k} dataset \footnote{When mentioning the \texttt{ImageNet} dataset throughout the paper, we refer to the \texttt{ImageNet-1k} version.} \citep{ImageNet} (trained on non-radio data), shown with black dots in Fig.~\ref{fig:rgz_class_eval}. We found that self-supervised pre-trained models reach approximately 7-12\% better scores with respect to the baseline. Another valuable indication is that the two-step pre-training approach done for the \texttt{smart\_hulk\_smgps} model training provide better results compared to training over random or source-centred images alone. The improvement is, however, not very significant with respect to the \texttt{hulk\_smgps}, likely due to both the limited size of the \texttt{banner\_smgps} dataset and the absence of Galactic-like diffuse and extended sources in the RGZ dataset. By construction, we expect that the \texttt{banner\_smgps} model should be more specialized for this kind of source morphologies. This will be tested in a future analysis once we finalize a new test dataset with diffuse sources taken from ASKAP EMU observations.\\In Fig.~\ref{fig:rgz_confusion_matrix} we report the confusion matrix obtained over the RGZ test sample with a \texttt{hulk\_smgps} pre-trained and frozen backbone. The obtained misclassification rates suggest that a considerable fraction (10\% to 20\%) of sources, particularly those with two or three components, may be hard to be correctly distinguished from other classes. After a visual inspection of the misclassified sources, we found that in some cases the misclassification is rather due to dataset mislabelling, e.g. the ground truth label present in the dataset is not correct and the model is indeed predicting the expected class. Some examples are reported in Fig.~\ref{fig:rgz_misclassified_sources}. Future analysis should therefore take into consideration a revision of the RGZ dataset annotation.

\begin{figure}[htb]
\centering%
\includegraphics[scale=0.42]{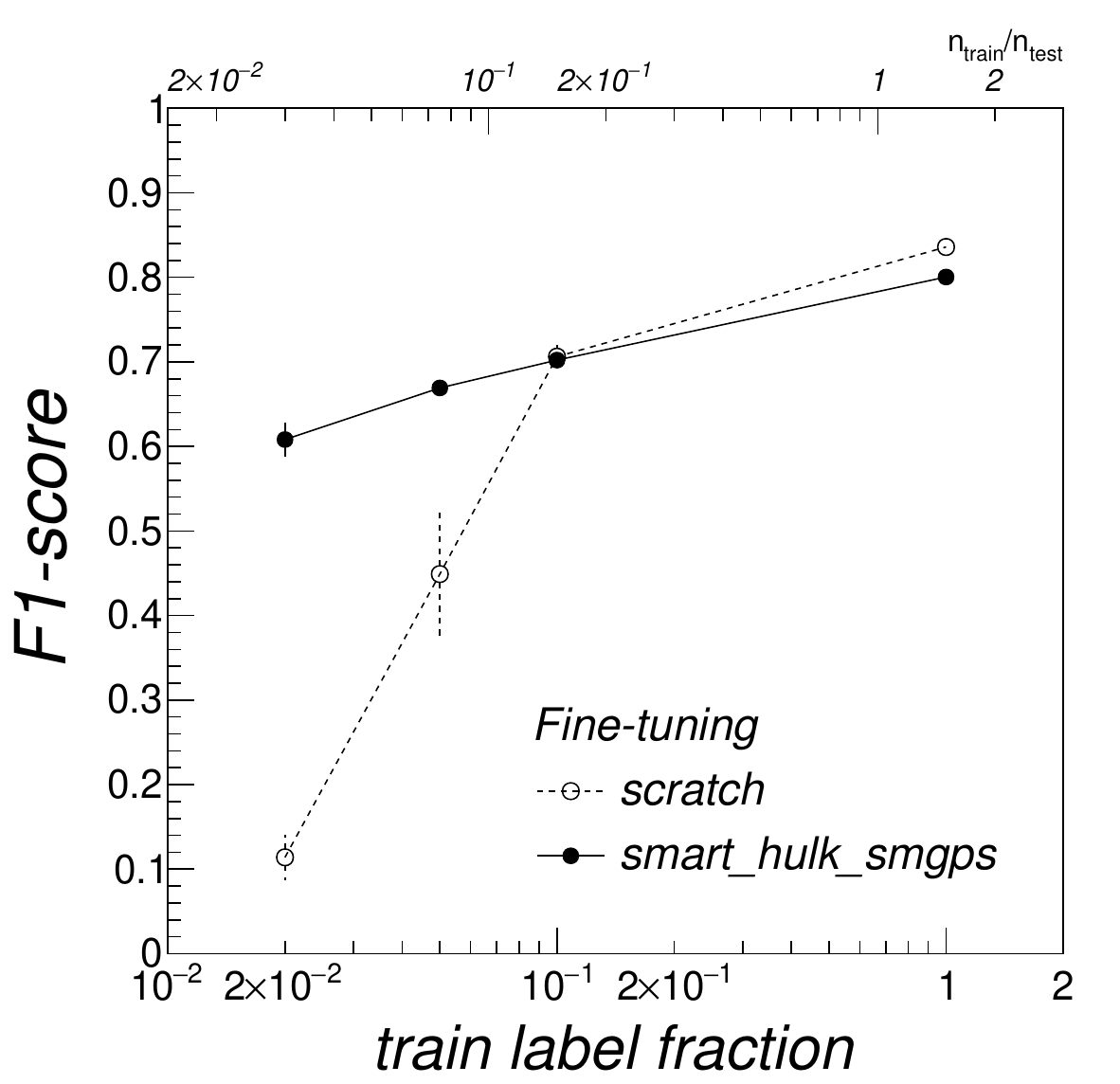}%
\caption{Classification F1-scores obtained (for all classes cumulatively) over RGZ test sets as a function of the train set size with two alternative models: one trained from scratch (open black dots), the other trained with backbone weights initialized to \texttt{smart\_hulk\_smgps} weights (filled black dots).}%
\label{fig:rgz_class_finetuning}
\end{figure}

\begin{figure*}[htb]
\centering%
\subtable[]{\includegraphics[scale=0.615]{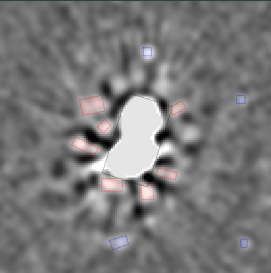}\label{fig:mrcnn_dataset_1}}%
\hspace{0.1cm}
\subtable[]{\includegraphics[scale=0.615]{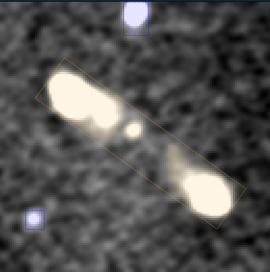}\label{fig:mrcnn_dataset_2}}%
\hspace{0.1cm}
\subtable[]{\includegraphics[scale=0.61]{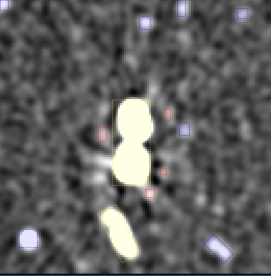}\label{fig:mrcnn_dataset_3}}%
\hspace{0.1cm}
\subtable[]{\includegraphics[scale=0.62]{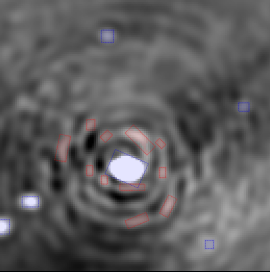}\label{fig:mrcnn_dataset_4}}%
\caption{Sample images (taken from \citealt{RiggiMaskRCNN}) from the dataset used for \emph{caesar-mrcnn} training/testing, including objects of different classes: a \texttt{FLAGGED} object (Fig.~\ref{fig:mrcnn_dataset_1}, in gray), \texttt{COMPACT} objects (in blue), a \texttt{MULTI-ISLAND} object (Fig.~\ref{fig:mrcnn_dataset_2}, in orange), \texttt{EXTENDED} objects (Fig.~\ref{fig:mrcnn_dataset_3}, in yellow), \texttt{SPURIOUS} objects (Fig.~\ref{fig:mrcnn_dataset_4}, in red).}%
\label{fig:mrcnn_dataset}
\end{figure*}

\subsection{Model fine-tuning}
\label{subsec:rgz-finetuning}
We fine-tuned the source classifier by unfreezing backbone layers (e.g. training them along with the classification head) and compared the accuracies reached by two models: one initialized with random weights (e.g. training from \emph{scratch}), and the other with backbone weights initialized to the \texttt{smart\_hulk\_smgps} backbone weights (best performing model found in Section~\ref{subsec:rgz-eval}). 
We compared the results of both models when trained on the full training sets and when trained on smaller training sets, obtained by gradually removing labelled data randomly from the original set. In all cases, models were trained for 150 epochs.
The test sets were kept unchanged to compute the classification accuracies. This was done to study how the model performs in the recurring scenario in which the amount of labelled data is significantly limited. 
We reported the results in Fig.~\ref{fig:rgz_class_finetuning}. As can be seen, the fully supervised model (trained from scratch) becomes almost untrainable, providing poor classification metrics, in the small number of labels regime. This occurs for the RGZ dataset below a fraction of approximately 10\% of the original train dataset. On the other hand, self-supervised pre-training enables to fine-tune the model even with few labels, achieving considerably better metrics (>20\%).
Above the 10\% label fraction threshold, the fully supervised model achieved slightly better scores, highlighting that no significant performance benefits are obtained from the pre-training process, at least with the model and dataset sample sizes available for this work.

\section{Task II: Radio source detection}
\label{sec:source-detection}
In this section, we quantitatively evaluate the learned self-supervised representation on an instance segmentation problem, specifically the detection of radio sources with various morphologies.\\Algorithms used in traditional radio source finders are not well-suited for detecting extended radio sources with diffuse edges, and they are unable to detect extended sources that are composed of multiple disjoint regions. To address this limitation, new source finders \citep{Wu2019,Mostert2022,Zhang2022,Yu2022,RiggiMaskRCNN,Lao2023,Gupta2024b,Cornu2024} based on deep neural networks and object detection frameworks have been developed and trained on either simulated or real radio data. Core components of these models are deep CNN backbones and transformer architectures, both of which have millions of parameters that need to be optimized during training. 
Although these models offer a substantial advancement in extended radio galaxy detection, their performance is limited by the small size (few thousand images) and the imbalance of objects in the available radio training datasets. Additionally, there is a potential performance drop (up to 10\% in \citealt{RiggiMaskRCNN}) when transferring a trained model to data produced by a different survey or telescope, especially if the new data has a better angular resolution \citep{Tang2019}. 
To improve the training stage, it is a common practice to use models pre-trained on much larger annotated samples of non-astronomical images, such as the \textit{ImageNet-1k} (\citealt{ImageNet}, $\sim$1.28 million images) or the \textit{COCO} (\citealt{COCO}, 328,000 images) datasets. 
In this scenario, it is worth exploring whether foundational models built with self-supervised methods on unlabelled radio data can offer performance benefits over non-radio foundational models, especially with small datasets.

\subsection{\textit{caesar-mrcnn} source detector}
For this analysis, we used the \emph{caesar-mrcnn} source detector \citep{RiggiMaskRCNN}, based on the Mask R-CNN object detection framework \citep{MaskRCNN}, to extract source segmentation masks and predicted class labels from input radio images. With respect to our original work, we have upgraded the software to TensorFlow 2.x, producing a new refactored version\footnote{\url{https://github.com/SKA-INAF/caesar-mrcnn-tf2}} with an improved data pre-processing pipeline and support for additional backbone models.\\ 
In this context, we would like to make a brief preamble and clarify the motivations that guided the development of the \emph{caesar-mrcnn} source detector, as these were either misinterpreted or inaccurately presented in other works. Additionally, we aim to address certain conceptual aspects that we realize are often source of confusion within this field.\\It is essential to recognize that source detection (or extraction), classification and source characterization (or measurement) represent distinct conceptual stages. A source detector, to be defined as such, should focus solely on extracting source bounding boxes or, preferably, pixel masks, which are the inputs required for the source measurement or classification stages. The source measurement step, on the other hand, is responsible for estimating source properties such as position, flux density, and shape from the outputs of the source detection. Strictly speaking, this step is not required in a source detector, as assumed in \cite{Lao2023}. From a methodological standpoint, it is advisable to avoid conflating these stages. This may allow addressing numerous use cases simultaneously, but it can also be counterproductive, leading, for example, to design compromises and overly complex models with multiple loss components to be balanced during training. The resulting models likely have a higher chance of underperforming on both tasks (detection or characterization) with respect to models that are designed and optimised for a specific task. For this reason, source characterization metrics should be independently evaluated and not mixed with the detection metrics, as required, for example, in the SKA Data Challenge 1 \citep{Bonaldi2021} scoring function. When we designed the \emph{caesar-mrcnn} source detector, we deliberately did not provide a source characterization stage. As we already implemented source measurement functions in the \emph{caesar} source finder, we rather aim to interface both codes and, at best, add new developments for improvements in specific areas, such as low S/N source characterization and source deblending, as discussed in \cite{Boyce2023}.\\In recent ML-based source extractors, source classification was typically performed alongside the detection step, often to classify extracted sources into compact and extended classes of radio galaxies (FR-I, FR-II, etc.).
We aimed for our source detector to be general-purpose, portable, and not tied to a specific radioastronomical domain. Therefore, in our view, the detection step should, at a minimum, classify between real and spurious sources, or, at most, between domain-agnostic morphological classes. 
More refined or domain-specific classification schemes can be more effectively applied afterwards in specialized classifiers working on source-centered images obtained from the detection step. These considerations were the rationale behind the general class labeling scheme adopted in \emph{caesar-mrcnn} (briefly reported in the following Section).

\begin{figure}[htb]
\centering%
\includegraphics[scale=0.42]{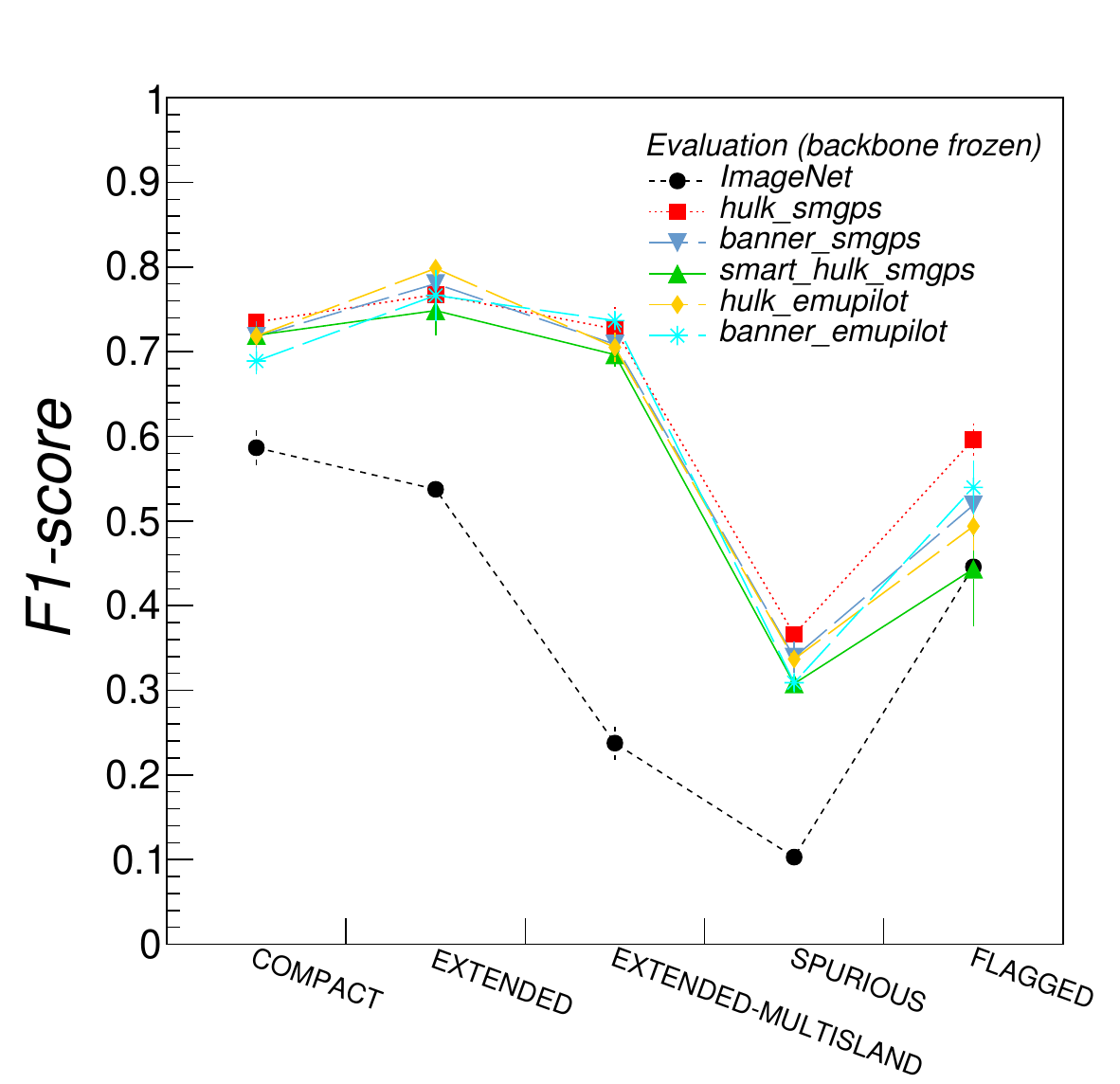}%
\caption{Mask R-CNN object detection F1-score metric obtained for different object classes over multiple test sets with different pre-trained and frozen backbones: \texttt{hulk\_smgps} (red squares), \texttt{banner\_smgps} (blue iverted triangles), \texttt{smart\_hulk\_smgps} (green triangles), \texttt{hulk\_emupilot} (orange diamonds), \texttt{banner\_emupilot} (cyan asterisks), \texttt{ImageNet} (black dots). The reported values and errors are the means and mean errors computed over 5 test sets.}%
\label{fig:mrcnn_eval}
\end{figure}

\subsection{Dataset}
To train and test \emph{caesar-mrcnn}, we considered the dataset described in \cite{RiggiMaskRCNN}, which contains 12,774 annotated radio images from different surveys, including VLA FIRST, ATCA Scorpio \citep{Umana2015}, and ASKAP-EMU Scorpio \citep{Umana2021}. The annotation data consist of pixel segmentation masks and classification labels for a total of 38,342 objects (both real and spurious sources) present in the dataset images. Five object classes were defined: 
\begin{itemize}
\item \texttt{SPURIOUS}: imaging artefacts around bright sources, having a ring-like or elongated compact morphology;
\item \texttt{COMPACT}: single-island isolated point- or slightly resolved compact radio sources with one or more blended components, each with morphology similar to the synthesized beam shape;
\item \texttt{EXTENDED}: radio sources with a single-island extended morphology, with one or more blended components, some morphologically different from the synthesized beam shape;
\item \texttt{EXTENDED-MULTISLAND}: radio sources with an extended morphology, consisting of more than one island, each eventually containing one or more blended components, having a point-like or an extended morphology; 
\item \texttt{FLAGGED}: poorly-imaged single-island radio sources, highly contaminated by nearby imaging artefacts.
\end{itemize}
For more details on the dataset labelling schema and rationale, we refer the reader to the original work. We also define a generic class label \texttt{SOURCE} for analysis purposes, including real and non-flagged sources, i.e. object instances of class \texttt{COMPACT}, \texttt{EXTENDED}, or \texttt{EXTENDED-MULTISLAND}. Though it is planned, the dataset does not presently contain images and annotation data for Galactic diffuse objects. Indeed, none of existing ML-based finders have been trained to detect diffuse sources other than radio galaxy diffuse structures (e.g. lobe components). The latter are the only diffuse structures present in our dataset, but we never noted to obtain poor detection performances on them, as reported in \cite{Ndungu2023}.\\In Fig.~\ref{fig:mrcnn_dataset} we present sample images from the dataset, including representative sources for each class. Given that the current dataset is significantly skewed towards compact sources (comprising approximately 80\% of the annotated objects), we created five re-balanced training samples, each containing 3245 images, with the following class distributions: \texttt{SPURIOUS} (1464 objects, 14.4\%), \texttt{COMPACT} (5457 objects, 53.6\%), \texttt{EXTENDED} (2042 objects, 20.1\%), \texttt{EXTENDED-MULTISLAND} (1047, 10.3\%), \texttt{FLAGGED} (169 objects, 1.7\%). The remaining data was reserved to create five test samples, each containing 5110 images, with the following class distributions: \texttt{SPURIOUS} (1022 objects, 6.6\%), \texttt{COMPACT} (12.346 objects, 80.0\%), \texttt{EXTENDED} (1307 objects, 8.5\%), \texttt{EXTENDED-MULTISLAND} (636, 4.1\%), \texttt{FLAGGED} (122 objects, 0.8\%).

\begin{figure}[htb]
\centering%
\includegraphics[scale=0.42]{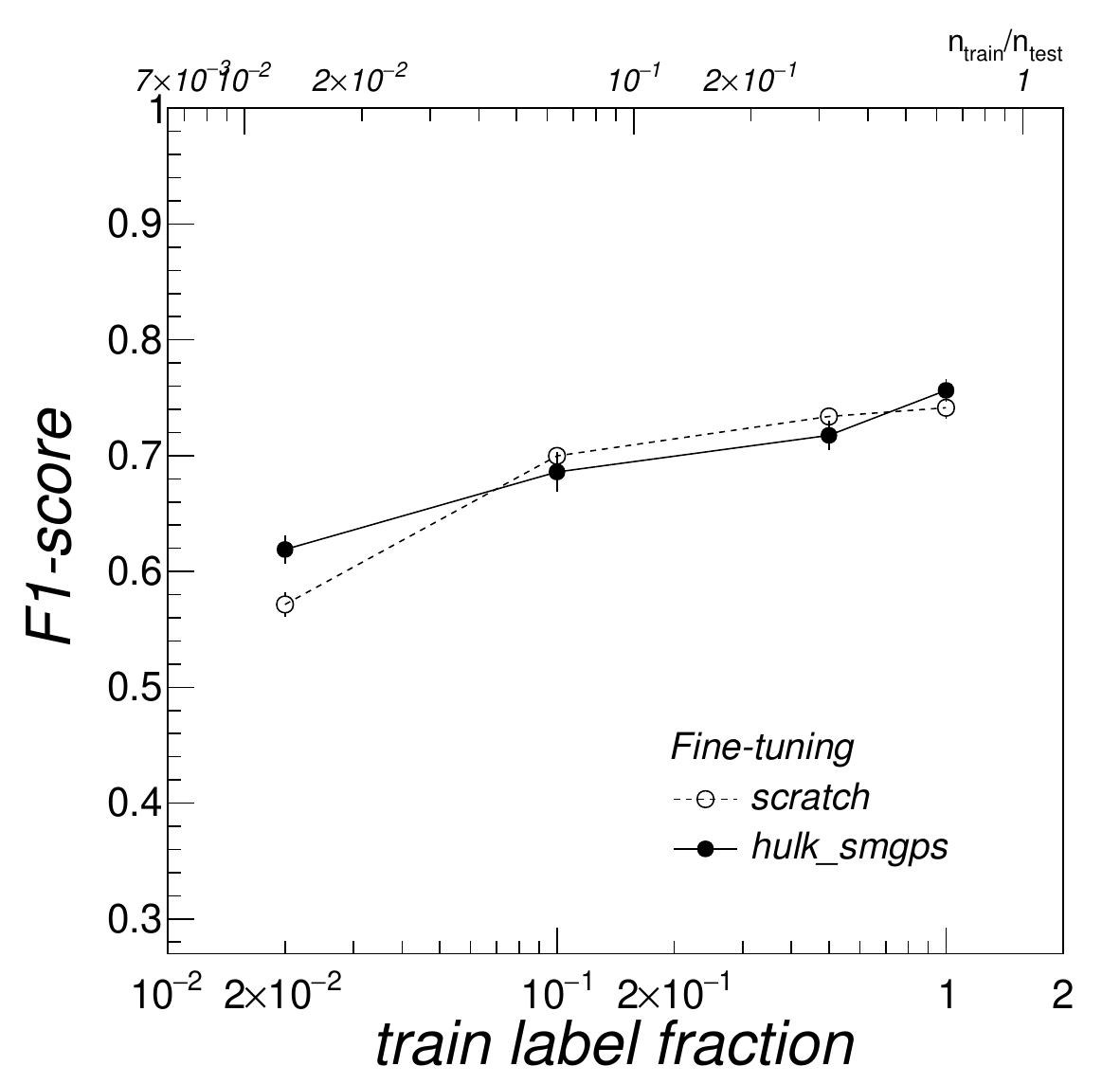}%
\caption{Mask R-CNN object detection F1-score metric obtained over the \texttt{SOURCE} class over multiple test sets as a function of the train set size with two alternative models: one trained from scratch (open markers), the other trained with backbone weights initialized to \texttt{hulk\_smgps} weights (filled markers).}%
\label{fig:mrcnn_finetuning}
\end{figure}

\subsection{Evaluation of self-supervised representation}
To assess the effectiveness of the self-supervised representation, we followed the procedure outlined in Section~\ref{subsec:rgz-eval}. We froze the Mask R-CNN model's \emph{ResNet18} backbone, setting and keeping its weights fixed to those obtained in the trained SimCLR models, and trained the remaining components (region proposal network, classification and bounding box regression head, mask prediction head) on multiple training datasets for a set number of epochs (250). 
The parameters of Mask R-CNN were configured to match the values optimized in our previous work (refer to \citealt{RiggiMaskRCNN}, Table A1). We applied the same pre-processing transformations used for training the self-supervised models (as detailed in Section~\ref{subsec:simclr-preproc}). During training, we applied three distinct image augmentations: rotation, horizontal/vertical flipping, and zscale transformation with random contrast in the range of 0.25 to 0.4.\\
The performance of source detection was evaluated on the test sets using the following metrics,
computed with a 0.5 object detection score threshold and an Intersection-over-Union (IoU) match threshold of 0.6 between detected and ground truth object masks:
\begin{itemize}
\item \emph{Completeness} ($\mathcal{C}$): Fraction of true sources detected by Mask R-CNN and classified as belonging to the \texttt{SOURCE} class group;
\item \emph{Reliability} ($\mathcal{R}$): Fraction of detected objects classified as \texttt{SOURCE}  that indeed match to true sources;
\item \emph{F1-score}: the harmonic mean of completeness and reliability:
\begin{equation}
\text{F1-score}=2\times\frac{\mathcal{C}\times\mathcal{R}}{\mathcal{C}+\mathcal{R}}
\end{equation}
\end{itemize}
As the focus of this analysis is on source detection, we have used the naming convention typically adopted in astronomical source catalogue works, although these metrics are sometimes referred to as recall/precision in other studies. The above metrics were computed for each class label and reported in Fig.~\ref{fig:mrcnn_eval} for different models trained with frozen self-supervised backbone weights: \texttt{hulk\_smgps} (red squares), \texttt{banner\_smgps} (blue triangles), \texttt{smart\_hulk\_smgps} (green triangles), \texttt{hulk\_emupilot} (orange diamonds). Metrics obtained with frozen \textit{ImageNet} weights are shown with black dots. The performance boost obtained with self-supervised models is significant, around 15\%-20\% for most classes, and even larger for multi-island sources and imaging artefacts. This is somehow expected, given that these structures are not present in the \textit{ImageNet} dataset. Overall, for the source class group we are interested in, we did not notice significant differences among trained self-supervised models, after taking into account the run-to-run statistical uncertainties on the obtained metrics. We will therefore consider a representative model (\texttt{hulk\_smgps}) in the following fine-tuning analysis.

\subsection{Model fine-tuning}
We fine-tuned the Mask R-CNN model using random initialization weights (training from \textit{scratch}) and weights initialized to \texttt{hulk\_smgps} self-supervised model. 
We computed the object detection metrics over the source class group as a function of the training sample size, following the same approach discussed in Section~\ref{subsec:rgz-finetuning}. Results are reported in Fig.~\ref{fig:mrcnn_finetuning}. Black filled dots are the F1-scores obtained with the pre-trained \texttt{hulk\_smgps} model, while open black dots are those found when training from scratch. 
In this case, we did not observe a significant benefit from using self-supervised pre-training compared to the source classification task studied in Section~\ref{sec:morph_classification}. The improvement in performance in the low label regime (<10\% of the original training sample size) is, in fact, of the order of a few percent. This behaviour highlights that other Mask R-CNN components likely play a major role in the overall model detection performance with respect to the backbone network. 

\begin{figure*}[htb]
\centering%
\captionsetup[subfigure]{labelformat=empty}
\subtable[\scriptsize{\texttt{\{RADIO-GALAXY,DIFFUSE\}}}]{\includegraphics[scale=0.27]{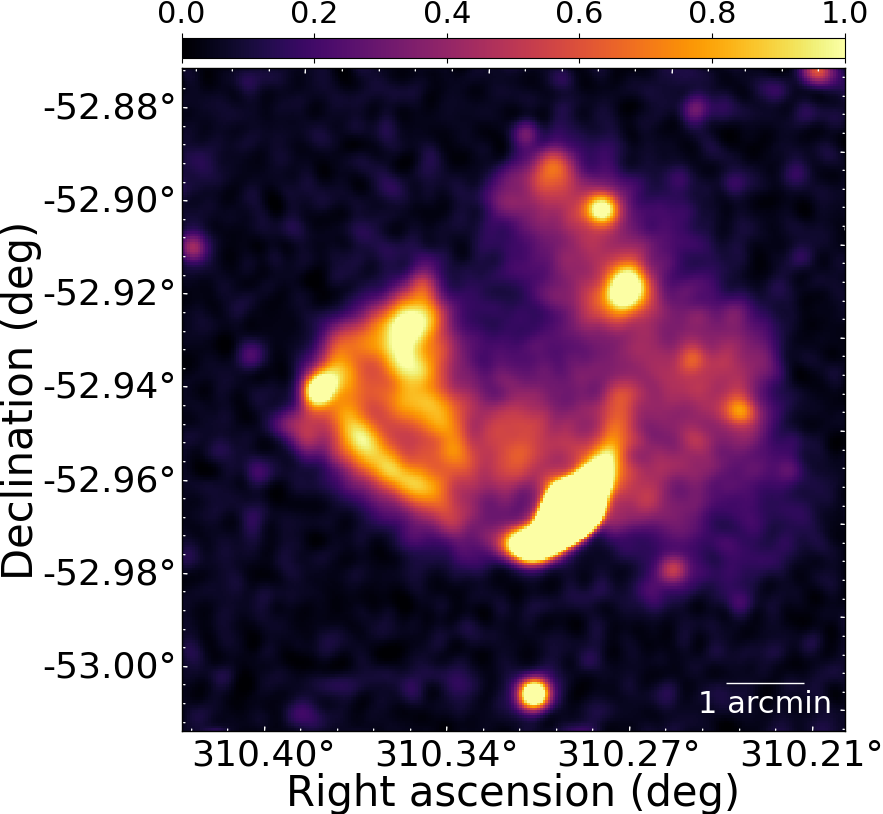}\label{fig:hulk_emupilot_examples_1}}%
\subtable[\scriptsize{\texttt{\{RADIO-GALAXY,ARTEFACT\}}}]{\includegraphics[scale=0.27]{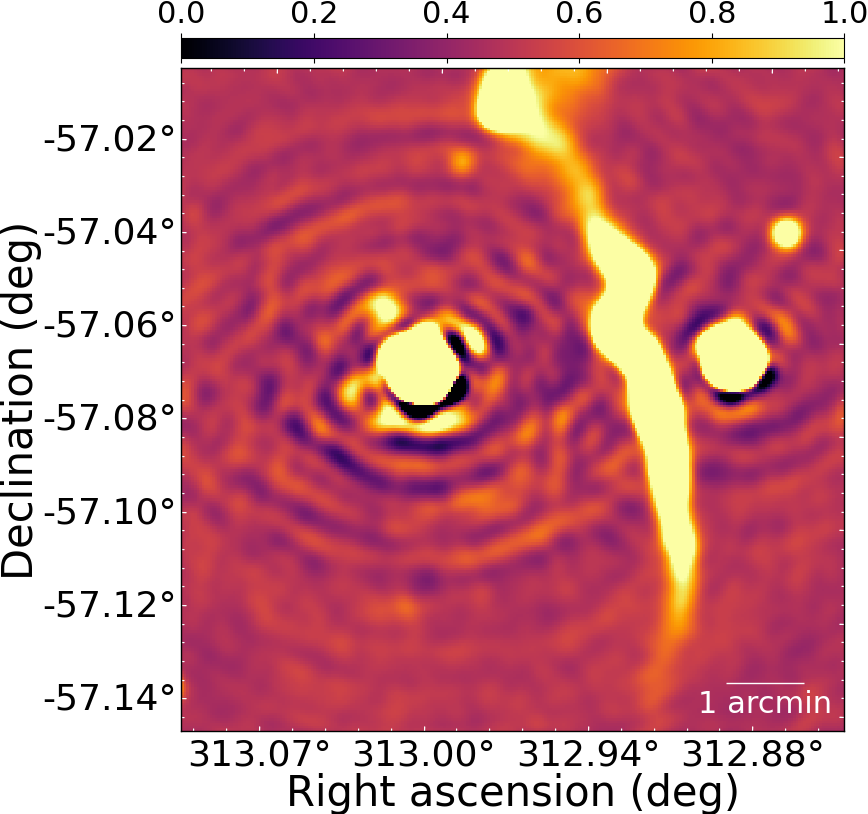}\label{fig:hulk_emupilot_examples_2}}%
\subtable[\scriptsize{\texttt{\{DIFFUSE-LARGE,BORDER\}}}]{\includegraphics[scale=0.27]{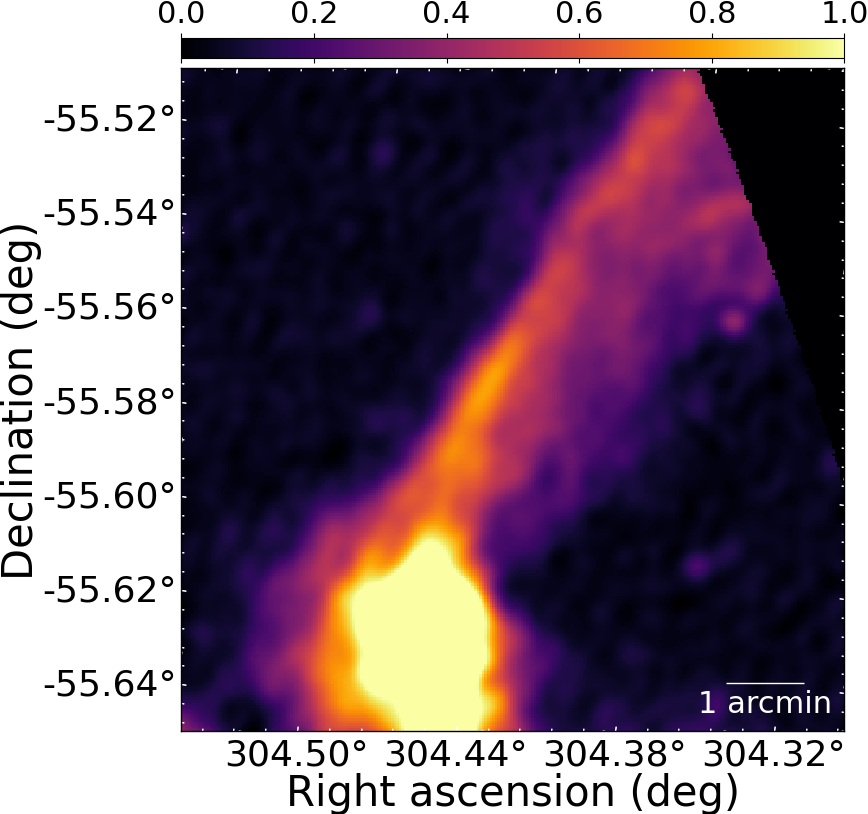}\label{fig:hulk_emupilot_examples_3}}\\%
\subtable[\scriptsize{\texttt{\{RADIO-GALAXY\}}}]{\includegraphics[scale=0.27]{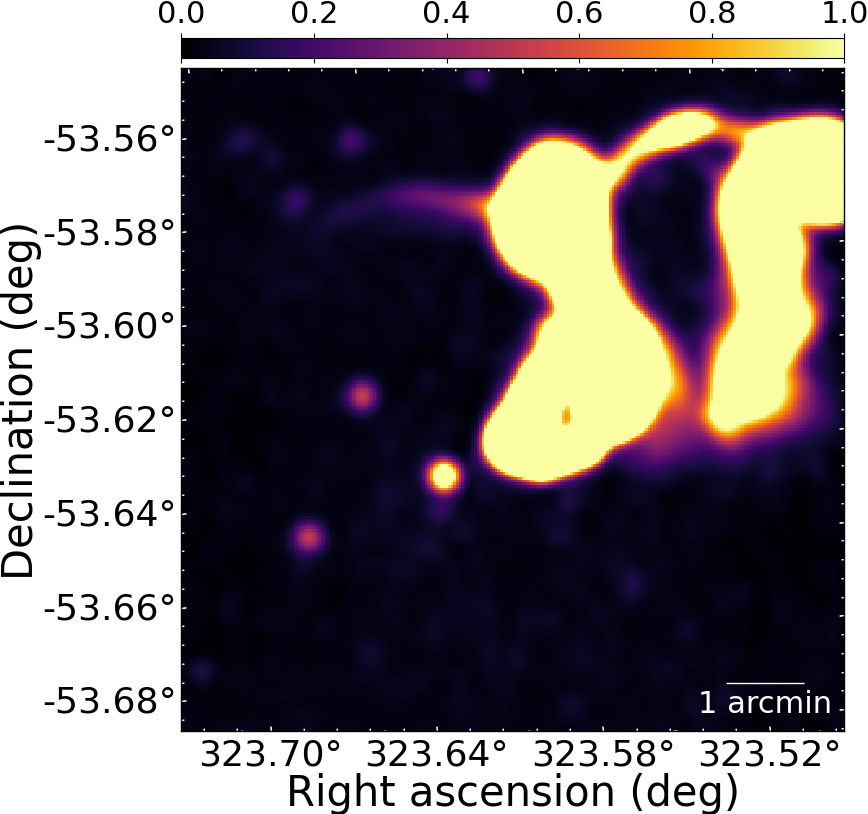}\label{fig:hulk_emupilot_examples_4}}%
\subtable[\scriptsize{\texttt{\{DIFFUSE-LARGE\}}}]{\includegraphics[scale=0.27]{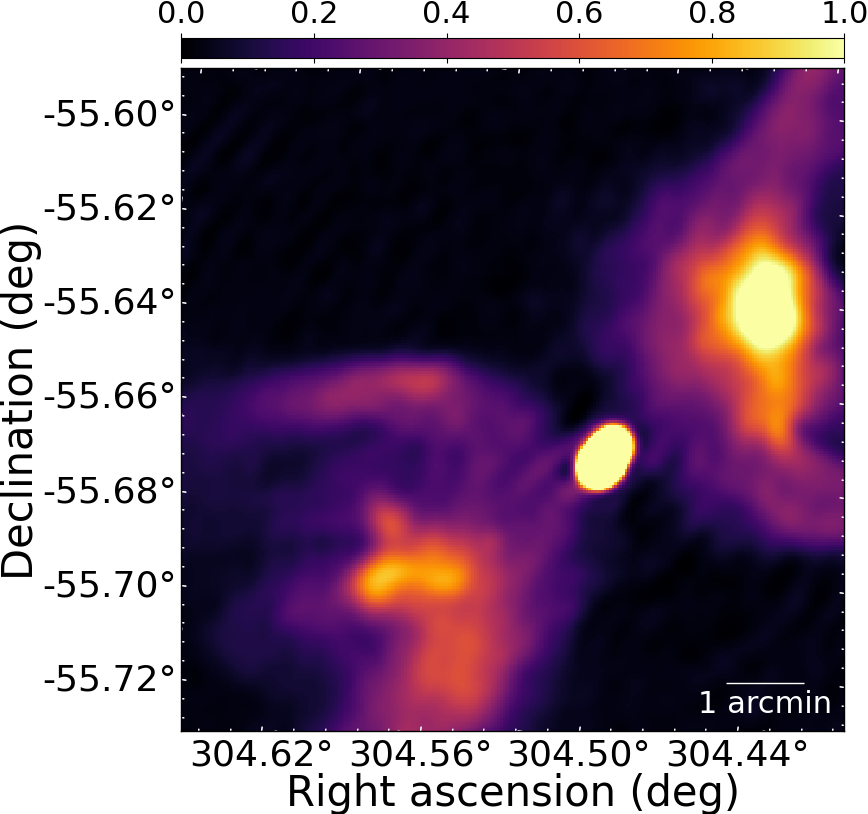}\label{fig:hulk_emupilot_examples_5}}%
\subtable[\scriptsize{\texttt{\{RADIO-GALAXY\}}}]{\includegraphics[scale=0.27]{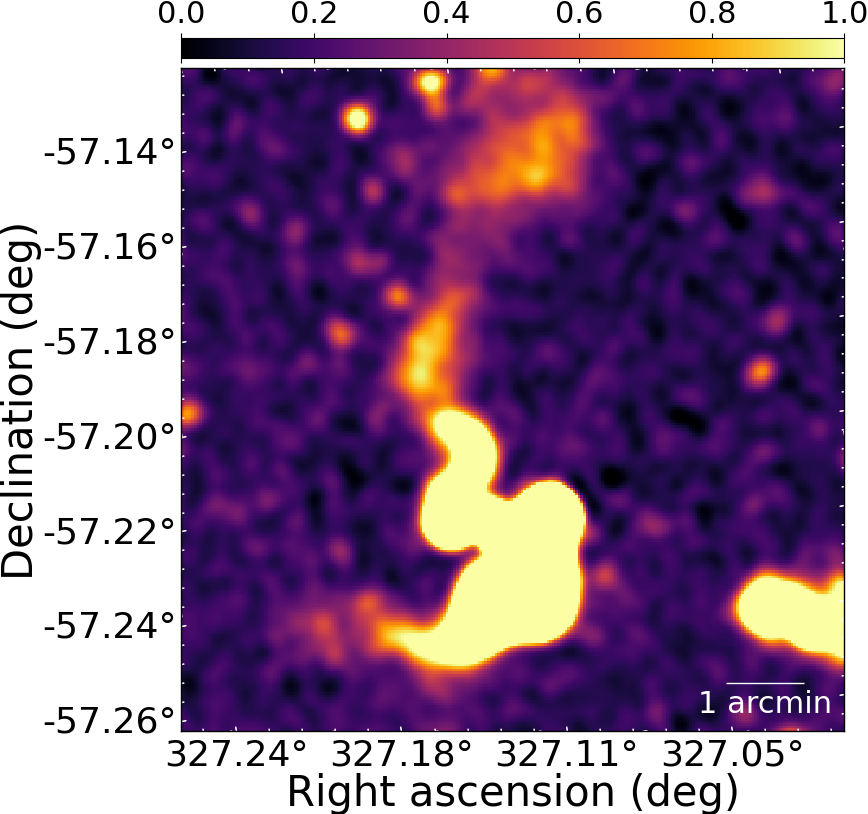}\label{fig:hulk_emupilot_examples_6}}%
\caption{Sample images from the \texttt{hulk\_emupilot} dataset, labelled as \texttt{PECULIAR} and \texttt{COMPACT}. The other assigned labels are reported below each frame. A zscale transform was applied to all images for visualization scopes.}%
\label{fig:hulk_emupilot_examples}
\end{figure*}

\section{Task III: Search for peculiar objects}
\label{sec:anomaly}
In this section, we quantitatively evaluated the learned self-supervised representations in an anomaly detection problem, i.e. employing them for an unsupervised search of radio objects with peculiar morphologies.\\Next-generation radio surveys carried out with SKA precursor telescopes are already generating a huge amount of data. Serendipitous discoveries were already reported and obtained by visual inspection of the observed maps. For instance, \cite{Norris2021b} and \cite{Koribalski2021} discovered a class of diffuse objects with a roundish shape, dubbed \emph{Odd Radio Circles} (ORCs), in the ASKAP EMU pilot survey, that did not correspond to any types of object or artifacts known to have similar morphological features. As it is extremely likely that new discoveries are still waiting to be found in such data deluge, astronomers have started to explore ML-based methods to automatically search for objects with peculiar morphologies. In this process, various methods were proposed, allowing to rediscover previously identified anomalies (including the first detected ORCs) and identify completely new objects \citep{Gupta2022,Lochner2023}.\\In this context, two major methodologies were used. \cite{Gupta2022} and \cite{Mostert2021} employed rotation and flipping invariant self-organizing maps (SOMs) to search for anomalies in the ASKAP EMU pilot and LOFAR LoTSS survey data, respectively. Both analysis used images of fixed size (approximately 1' to 5', $\sim$150 pixels per size), centered around previously catalogued radio sources. The Euclidean distance from each "representative" image in the SOM lattice was used as an “anomaly proxy”, e.g. anomalous images have larger Euclidean distances from their closest SOM template image.\\
\cite{Segal2023} used a coarse-grained complexity metric as an "anomaly" proxy to detect peculiar objects in the ASKAP EMU pilot survey. Their method is based on the idea that image frames containing complex and anomalous objects have a higher Kolmogorov complexity compared to ordinary frames. 
In contrast to the previously mentioned methods, \cite{Segal2023} conducted a blind search by sliding fixed image frames of size 256$\times$256 pixels ($\sim$12 arcmin) through the entire map, rather than focusing on frames centered around known source positions. An approximated complexity estimation for each frame was then computed from the compression file size (using the gzip algorithm) of smoothed and resized frames. This allowed the authors to obtain a catalogue of peculiar sources at different reliability levels, corresponding to different complexity threshold choices. The complexity metric is conceptually simple and fast to compute, which is undoubtedly a positive aspect of this method. However, as noted by \cite{Mostert2021}, the complexity metric may not fully capture the morphological features of the sources present in the images.\\
A potential limitation of "source-centric" approaches could be their reliance on catalogues created with traditional source finding algorithms, which are known to have a higher likelihood of missing diffuse sources (a primary target in anomaly searches). Nevertheless, existing studies successfully manage to identify new anomalous sources in their datasets. \cite{Mostert2021} also noted that their method is not fully sensitive to detect anomalies at angular scales much smaller than the chosen image size (100 arcsec in their work). The choice of the frame size is an aspect that certainly affects "blind" anomaly searches as well.
\\In this work, we aim to carry out a blind anomaly search study using a different method, which relies on image features extracted by trained self-supervised models. Details on the dataset used and the methodology are provided in the following paragraphs.

\begin{figure*}[htb]
\centering%
\subtable{\includegraphics[scale=0.45]{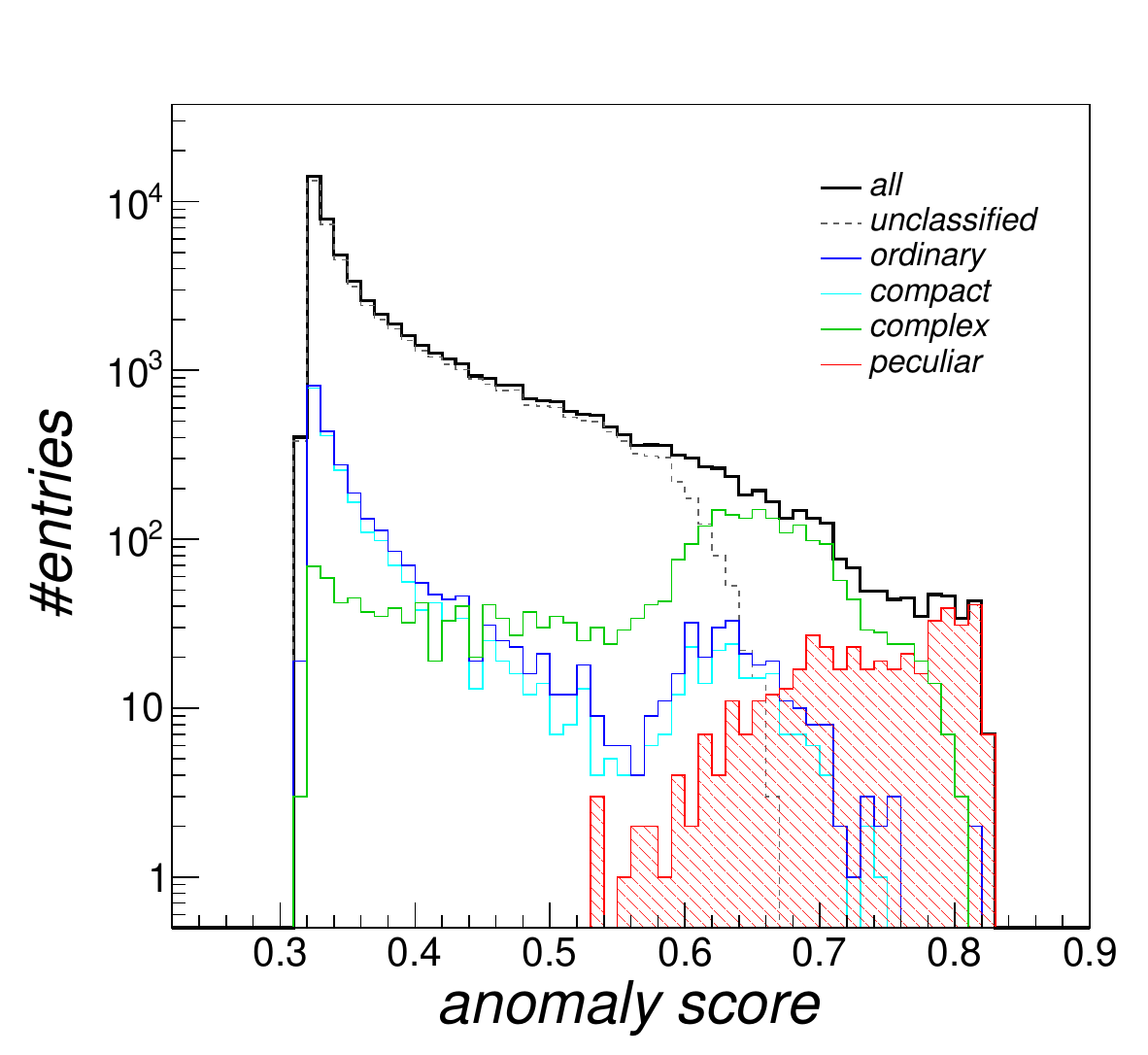}\label{fig:anomaly_metrics_1}}
\subtable{\includegraphics[scale=0.45]{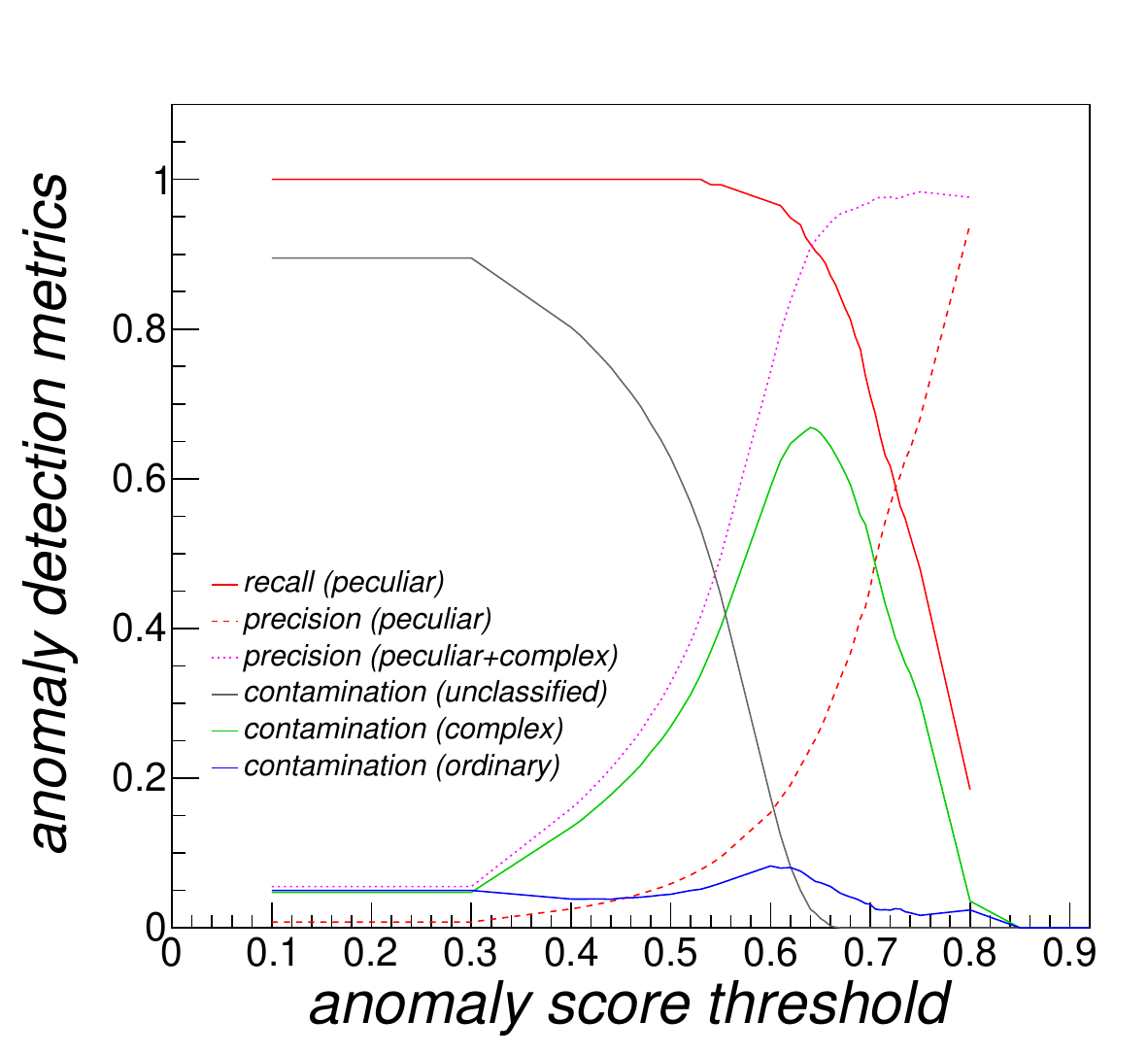}\label{fig:anomaly_metrics_2}}
\caption{Left: Anomaly score of frames contained in the \texttt{hulk\_emupilot} dataset, shown as black solid histogram, found with the \emph{Isolation Forest} algorithm over top-10 feature data. Unclassified frames are shown with a dashed line. Red filled histogram are the scores of peculiar frames. Ordinary frames (e.g. hosting only compact or artefacts) are shown in blue, pure compact frames in light blue, while frames not tagged as peculiar that host complex sources or structures (\texttt{EXTENDED}, \texttt{DIFFUSE}, \texttt{DIFFUSE-LARGE}, \texttt{RADIO-GALAXY}) are shown in green. Right: Anomaly detection metrics (recall, precision, contamination) as a function of the applied anomaly score threshold. Red solid and dashed lines indicate the recall and precision achieved on peculiar frame detection. Purple dotted line is the precision obtained over both peculiar and complex frames. The other solid coloured lines indicate the fraction of unclassified (black line), complex (green line) and ordinary frames contaminating the selected anomaly sample.}%
\label{fig:anomaly_metrics}
\end{figure*}

\subsection{Dataset}
\label{subsec:dataset_anomaly}
For this analysis, we considered the \texttt{hulk\_emupilot} dataset (55,774 images) described in Section~\ref{sec:ssc-dataset}. We annotated through visual inspection approximately 10\% of the data (5800 images) using the following set of labels: 
\begin{itemize}
\item \texttt{BACKGROUND}: If the image is purely background noise, e.g. no sources are visible. Typically, this label is set for frames located at the map borders;
\item \texttt{COMPACT}: if point sources or compact sources comparable with the synthesized beam size (say $<$10 times the beam) are present. Double or triple sources with point-like components also fall into this category;
\item \texttt{EXTENDED}: if any extended source is visible, e.g. a compact source with extension $>$10 $\times$ beam;
\item \texttt{RADIO-GALAXY}: if any extended source is visible with a single- or multi-island morphology, suggesting that of a radio galaxy (e.g. core + lobes);
\item \texttt{DIFFUSE}: if any diffuse source is visible, typically having small-scale (e.g. $<$few arcmin) and roundish morphology;
\item \texttt{DIFFUSE-LARGE}: if any large-scale (e.g. covering half of the image) diffuse object with irregular shape is visible;
\item \texttt{FILAMENT}: if any extended filamentary structures is visible;
\item \texttt{ARTEFACT}: if any ring-shaped or ray-like artefact is visible, e.g. typically around bright resolved sources;
\item \texttt{PECULIAR}: if any object is found with peculiar/anomalous morphology;
\item \texttt{MOSAICKING}: if any residual pattern of the mosaicking process used to produce the image is present.
\end{itemize}
More than one label can be assigned to each image, depending on the object/features the user recognize in the image.\\A total of 428 peculiar frames were selected through visual inspection starting from a list of 1198 peculiar frames identified in \cite{Segal2023} with a complexity metric analysis and from a catalogue of 361 peculiar sources reported in \cite{Gupta2024}. In Fig.~\ref{fig:hulk_emupilot_examples} we show examples of peculiar images from the dataset with their annotation labels.

\begin{table}[htb]
\centering%
\footnotesize%
\caption{Peculiar frame detection metrics obtained with the \emph{Isolation Forest} algorithm over selected feature sets (column (1)) when using an anomaly score threshold (reported in column (2)) that provides the best compromise in terms of peculiar frame recall and precision, respectively shown in columns (3) and (4). The precision relative to joint peculiar and complex frames is shown in column (5). The fractions of complex and ordinary frames contaminating the predicted anomalous sample are shown in columns (6) and (7).}
\begin{tabular}{lcccccc}
\hline%
\hline%
 &  & $\mathcal{R}$ & $\mathcal{P}$ & $\mathcal{P}_{pec+complex}$ (\%) & $\mathcal{C}_{complex}$ & $\mathcal{C}_{ordinary}$ \\%
\multirow{-2}{*}{Features} & \multirow{-2}{*}{Thr.} & (\%) & (\%) & (\%) & (\%) & (\%)\\%
\hline%
\texttt{top2} & 0.700 & 55.6 & 59.8 & 93.5 & 33.7 & 6.5\\%
\texttt{top5} & 0.750 & 61.2 & 63.0 & 93.0 & 30.0 & 7.0\\%
\texttt{top10} & 0.725 & 59.1 & 58.7 & 97.4 & 38.7 & 2.6\\%
\texttt{top15} & 0.660 & 57.7 & 58.7 & 95.5 & 36.8 & 4.5\\%
\hline%
\hline%
\end{tabular}
\label{tab:anomaly_results}
\end{table}

\subsection{Anomaly analysis}
\label{subsec:analysis_anomaly}
The data representation variables are each sensitive to different features of the images, including details (e.g. the presence of image borders or artifacts, background noise or mosaicking patterns, compact source density, etc) that are not relevant for the anomaly search task. We tried to limit the dependency on background features with the \emph{RandomThresholding} augmentation, but the model was not fully made invariant with respect to the other aspects. For this reason, we carried out a feature selection analysis, aiming to explore and select features that are mostly correlated with the presence of objects with diffuse or extended morphology. We divided the labelled set of images into two groups: "interesting" frames include images labelled as \{\texttt{EXTENDED},\texttt{DIFFUSE},\texttt{DIFFUSE-LARGE}\}, while "ordinary" frames include the rest of labelled images, mostly hosting only compact sources or artifacts around them. We then trained a LightGBM \footnote{LightGBM is a high-performance gradient boosting framework based on decision tree algorithm, particularly suited for classification tasks on tabular data. More details are available at \scriptsize{\url{https://lightgbm.readthedocs.io/en/latest/index.html}}.} \citep{Ke2017} classifier to classify the two groups with all representation features (512) as inputs. A subset of available data was reserved as a cross-validation set for model training early stop. Using shallow decision trees (\texttt{max\_depth}=2) and default LightGBM parameters (\texttt{num\_leaves}=32, \texttt{min\_data\_in\_leaf}=20), we obtained a classification score of 75.3\%. In Fig.~\ref{fig:lgbm-feat-importance} we report a plot showing the feature importance returned by the LightGBM trained model. As one can see, a small set of features are identified as the most powerful for selecting interesting frames. We therefore carried out the following data exploration and unsupervised analysis, restricting the parameter set to the top-15 ranked variables in importance.\\In Fig.~\ref{fig:hulk_umap_vs_pars_1} we report a two-dimensional projection of the top-15 variables produced with the \emph{Uniform Manifold Approximation and Projection} (UMAP) dimensionality reduction algorithm \citep{UMAP} as a function of the image noise rms level in logarithmic scale (coloured z-axis). As can be seen, the obtained representation shows a residual dependency on physical image parameters, such as the noise rms, that cannot be fully removed by the augmentation scheme currently adopted. In the other panels of Fig.~\ref{fig:hulk_umap_vs_pars} we report the same projection for unlabelled (gray markers) and labelled data, shown with coloured markers. Interestingly, frames that were labelled as peculiar or complex (e.g. containing extended/diffuse objects or artifacts) tend to cluster in specific areas of the projected feature space, also related with higher noise areas, while ordinary frames are uniformly spread in the feature space. Other higher noise areas present in Fig.~\ref{fig:hulk_umap_vs_pars_1} seem related to frames that are closer to the mosaic edges or having artifacts (see Fig.~\ref{fig:hulk_umap_vs_pars_2}).\\We searched for peculiar frames using the \emph{Isolation Forest} (IF) \citep{IsolationForest} outlier detection algorithm\footnote{\emph{Isolation Forest} is an unsupervised decision-tree-based algorithm for outlier detection in tabular data, that works by randomly selecting a feature and a random split value to isolate data points in a binary tree. It identifies outliers as instances that require fewer splits to be isolated, exploiting the inherent rarity of anomalies in a dataset.}. We tuned these IF hyperparameters using the annotated dataset:
\begin{itemize}
\item \texttt{contamination}: The proportion of outliers in the data set. We scanned these values: {'auto', 0.001, 0.01, 0.1}.
\item \texttt{max\_samples}: The number of samples to draw from the training data to train each base estimator. We scanned these values: {'auto', 0.001, 0.01, 0.02, 0.05, 0.1, 0.2, 0.3, 0.4, 0.5, 0.6, 0.7, 0.8, 0.9, 1.0}.
\end{itemize}
Scans were repeated for different choices of importance ranked feature sets: \texttt{top2}, \texttt{top5}, \texttt{top10}, and \texttt{top15}. A number of 200 base estimators were used in the tree ensemble. Other IF parameters were set to defaults. Best classification results were obtained with a smaller fraction of samples (\texttt{max\_samples}=0.02) and \texttt{contamination}=0.001.\\We then ran the IF algorithm in an unsupervised way with tuned parameters and obtained an anomaly score for each dataset frame. The anomaly score ranges from 0 to 1, with most anomalous data expected to have values close to 1. In Fig.~\ref{fig:anomaly_metrics} (left panel) we report the distribution of IF anomaly scores of all frames contained in the \texttt{hulk\_emupilot} dataset, shown as a black solid histogram, found over top-10 feature data. Unclassified frames are shown with a dashed line. The red filled histogram indicates the labelled peculiar frames. Ordinary frames (e.g. hosting only compact or artifacts) are shown in blue, pure compact frames in light blue, while complex frames (e.g. hosting extended or diffuse structures, not labelled as peculiar) are shown in green. We computed the following anomaly detection metrics as a function of the applied anomaly score threshold:
\begin{itemize}
\item \emph{Recall} ($\mathcal{R}$): Fraction of peculiar frames that were correctly detected by the model above the applied score threshold out of all frames labelled as peculiar;
\item \emph{Precision} ($\mathcal{P}$): Fraction of frames correctly classified as peculiar, out of all frames the model predicted to be peculiar, above the applied score threshold;
\item \emph{Contamination} ($\mathcal{C}$): Fraction of non-peculiar frames of a given label detected above the applied score threshold. 
\end{itemize}
Peculiar frame recall and precision are reported in Fig.~\ref{fig:anomaly_metrics} (right panel) as a function of the applied anomaly score threshold for top-10 feature data, respectively shown with solid and dashed red lines. We also computed the precision in classifying detected frames as either peculiar or complex, shown with a dotted purple line. The other solid coloured lines indicate the fraction of unclassified (black line), complex (green line) and ordinary frames contaminating the selected anomaly sample. In Table~\ref{tab:anomaly_results} we summarized the metrics obtained for different feature sets for the anomaly score threshold that provides the best peculiar recall/precision compromise (e.g. the score at which recall and precision curves cross in Fig.~\ref{fig:anomaly_metrics_2}). Best detection performances ($\sim$60\%) are obtained with the top-5 features, but the top-10 feature set currently provides the smallest contamination fraction of ordinary frames ($<$3\%). When considering both peculiar and complex frames, the precision increases to 97\%.

\subsection{Astronomer-in-the-loop}
\label{subsec:anomaly_active_learning}
It is worth to note that the source peculiarity concept is rather subjective and may depend on the scientific domain of interest. For instance, a fraction of complex frames may well be considered as truly peculiar in specific analysis, and, on the other hand, missed peculiar frames may be considered not as relevant in other contexts. For this reason, an additional "human-in-the-loop" processing stage has to be applied to our list of candidate anomalies to create a refined sample that better fits scientific needs.\\For the sake of demonstration, we integrated our dataset in the \texttt{astronomaly} package\footnote{\url{https://github.com/MichelleLochner/astronomaly }} \citep{Lochner2021}. This allowed us to run an active learning process from a web interface in which users can personalize and sort the list of anomalous frames on the basis of the computed score and also their expressed preferences, such as how peculiar a frame is judged on a scale of 1 to 5. A screeshot of the \texttt{astronomaly} UI for our dataset is shown in Fig.~\ref{fig:astronomaly}.\\We plan to integrate in the future the full pipeline (feature extraction, anomaly detection, active learning loop) as a supported application within the \emph{caesar-rest} service\footnote{\scriptsize{\url{https://github.com/SKA-INAF/caesar-rest}}} \citep{Riggi2021}, and extend the web UI with missing functionalities (e.g. image filtering/exporting, model importing, configuration options, etc). In this study, we limited ourselves to primarily quantify the ordinary frame rejection power that can be currently achieved with self-supervised features, as this will largely impact the time needed to visually inspect the anomaly candidates in human-in-the-loop approaches to form the final anomaly sample. 


\section{Summary}
\label{sec:summary}
In this study, we investigated the potential of self-supervised learning for analysing radio continuum image data produced by SKA precursors. Specifically, we have used the SimCLR contrastive learning framework to train deep network models on large sets of unlabelled images extracted from the ASKAP EMU pilot and SARAO MeerKAT GPS surveys, either randomly selected or centred around catalogued extended source positions. The trained encoder network, based on the \textit{ResNet18} architecture, was used as a feature extractor and fine-tuned for three distinct downstream tasks (source detection, morphology classification, and anomaly detection) over test datasets comprising thousands of annotated images from other radio surveys (VLA FIRST, ASKAP Early Science, ATCA Scorpio surveys). Notably, some of these test datasets were purposefully created for this work.\\All trained models, including both the source code and network weights, have been publicly released. These represent a first outcome of this work, as they can be viewed as prototypal radio foundational models, available to be used in future applications for multiple scopes:
\begin{itemize}
\item to extract feature parameters from new radio survey images and perform data inspection, unsupervised classification or outlier detection analysis (as demonstrated in Section~\ref{sec:anomaly});
\item to serve as pre-trained backbone components of more complex models designed for source classification, detection or other tasks (e.g. source property characterization), eventually refined over new labelled datasets (as demonstrated in Sections~\ref{sec:morph_classification} and \ref{sec:source-detection}).
\end{itemize}
The analyses we performed in this work attempted to address various open questions in this field, paving the way for future analyses: 
\begin{itemize}
\item Do we observe any advantages stemming from self-supervised models trained on easily constructed "random" survey datasets compared to costly-to-compile "source-centric" datasets?
\item How does self-supervised learning on radio data compare in performance to models pre-trained on extensive non-radio datasets, such as \textit{ImageNet}?
\item Is it feasible to enhance existing radio source detectors utilizing deep networks through radio self-supervised pre-training?
\end{itemize}
We found that using uncurated large collections of unlabelled radio images randomly extracted from SKA precursor surveys resulted in significantly improved performances (approximately 5\%) in both radio source detection and classification tasks, compared to curated (albeit smaller) image samples extracted around extended source catalogues. This indication, primarily attributed to the augmented number of accessible images achievable with uncurated collections, is highly encouraging, as it suggests that certain aspects of source analysis can be enhanced even without investing numerous work months in catalogue production.\\The advantages gained from self-supervised pre-training on radio data, compared to non-radio data, are notably significant (exceeding 10\%) in both source classification and detection tasks. However, when contrasting our findings with fully supervised models trained from scratch, we observed that these benefits are only relevant with small labelled datasets (on the order of a few hundred images). This is certainly a positive aspect, considering that many available annotated datasets (such as \textit{MiraBest} or similar radio galaxy classification datasets) typically fall within this size range. Nevertheless, in order to observe a substantial impact on larger datasets, it becomes imperative to improve both the self-supervised pre-training dataset and the model itself.
\\We have identified some areas of developments to be made in the near future to improve source analysis performance, and overcome the limitations encountered in this study. Firstly, we plan to increase the size of our pre-training \texttt{hulk} datasets, by leveraging the massive amount of unlabelled image data being delivered by large area surveys, such as ASKAP EMU, the Very Large Array Sky Survey (VLASS) \citep{VLASS}, or the LOFAR Two-metre Sky Survey (LoTSS) \citep{LOTSS} surveys. In this context, to reduce the computational load during training, it is crucial to explore effective and automated strategies for constructing semi-curated large-scale pre-training datasets, potentially comprising millions of images. This step may require the development of specialized algorithms to filter or weight image frames included in the pre-training dataset, aiming to maximize the balance between ordinary and complex objects "seen" by the model.\\Additionally, we have already started to train larger architectures and recent state-of-the-art self-supervised frameworks, particularly those based on Vision Transformers (ViTs), over the same datasets produced for this study. Results will be compared against the SimCLR baseline and presented in a forthcoming paper.



\begin{acknowledgement}
This scientific work uses data obtained from Inyarrimanha Ilgari Bundara / the Murchison Radio-astronomy Observatory. We acknowledge the Wajarri Yamaji People as the Traditional Owners and native title holders of the Observatory site. CSIRO’s ASKAP radio telescope is part of the Australia Telescope National Facility (\url{https://ror.org/05qajvd42}). Operation of ASKAP is funded by the Australian Government with support from the National Collaborative Research Infrastructure Strategy. ASKAP uses the resources of the Pawsey Supercomputing Research Centre. Establishment of ASKAP, Inyarrimanha Ilgari Bundara, the CSIRO Murchison Radio-astronomy Observatory and the Pawsey Supercomputing Research Centre are initiatives of the Australian Government, with support from the Government of Western Australia and the Science and Industry Endowment Fund.\\
This work made use of PLEIADI, a computing infrastructure installed and managed by INAF.
\end{acknowledgement}

\paragraph{Funding Statement}
This work received funding from the INAF CIRASA and SCIARADA projects.

\paragraph{Competing Interests}
None

\paragraph{Data Availability Statement}
\label{sec:data-availability}
The software code used in this work is publicly available under the GNU General Public License v3.0\footnote{\scriptsize{\url{https://www.gnu.org/licenses/gpl-3.0.html}}} on the GitHub repository \url{https://github.com/SKA-INAF/sclassifier/}. The trained model weights have been made available on Zenodo repository at \url{link-at-paper-acceptance}.

\printendnotes

\newpage%
\appendix%
\onecolumn%
\renewcommand{\thesection}{\Alph{section}}
\renewcommand\thefigure{\thesection.\arabic{figure}} 

\section{Supplementary plots}
\label{appendix:addon_plots}
\setcounter{figure}{0}

\begin{figure*}[htb]
\centering%
\subtable[\scriptsize{\texttt{1C-1P} (true), \texttt{1C-2P} (pred), \texttt{1C-2P} (true corr.)}]{\includegraphics[scale=0.25]{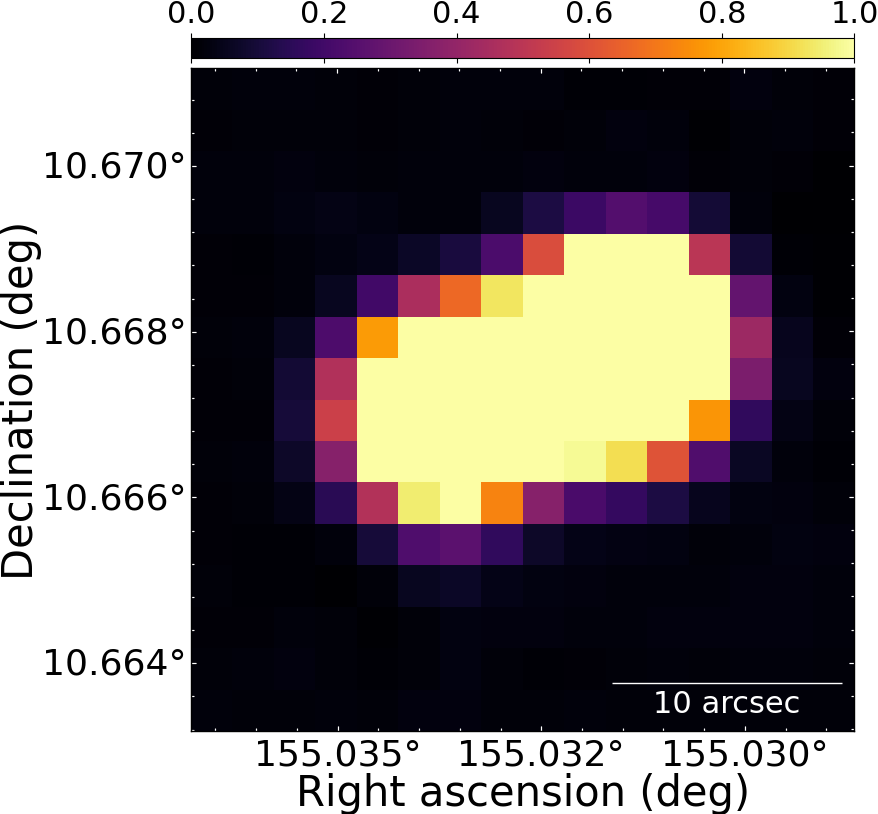}\label{fig:rgz_misclassified_sources_1}}
\subtable[\scriptsize{\texttt{1C-2P} (true), \texttt{1C-3P} (pred), \texttt{1C-3P} (true corr.)}]{\includegraphics[scale=0.25]{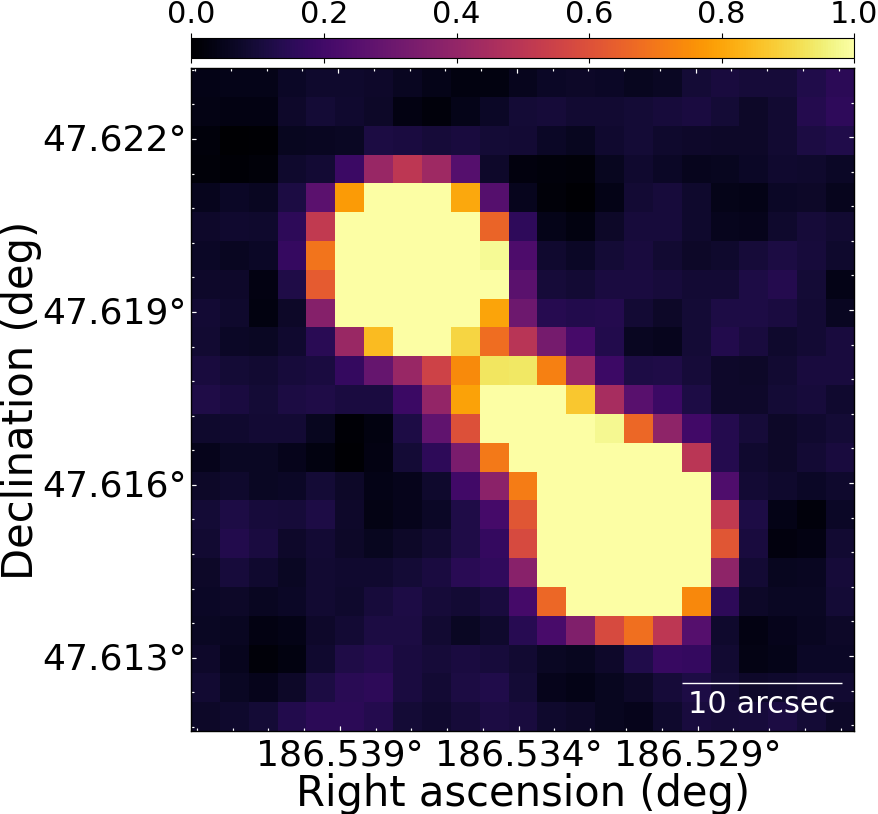}\label{fig:rgz_misclassified_sources_2}}
\subtable[\scriptsize{\texttt{1C-3P} (true), \texttt{1C-2P} (pred), \texttt{1C-2P} (true corr.)}]{\includegraphics[scale=0.25]{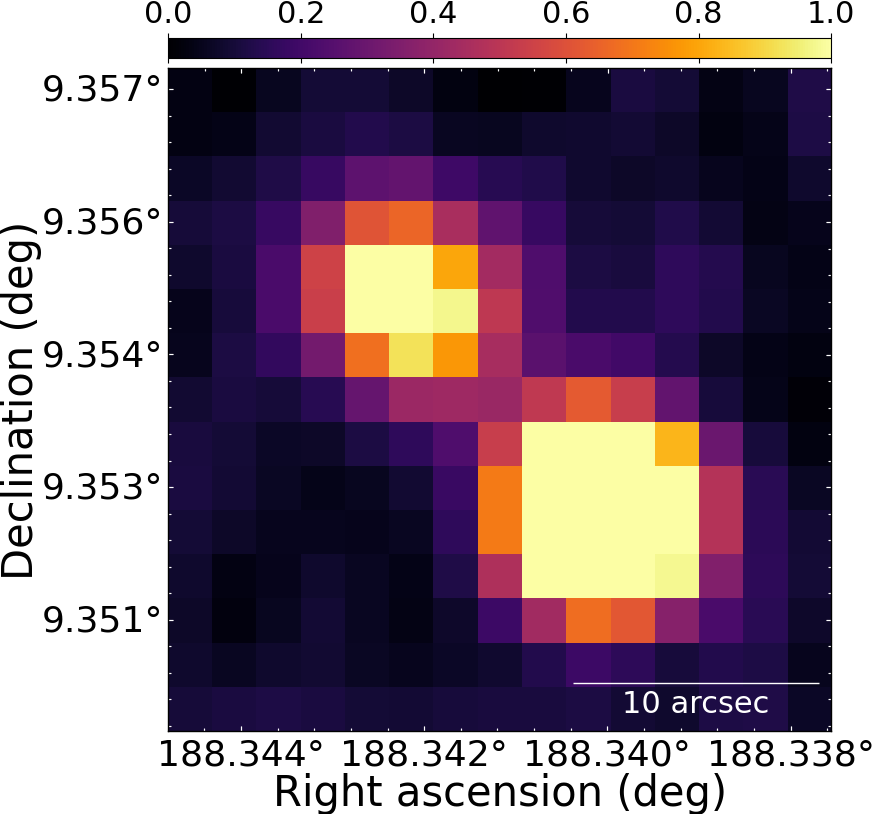}\label{fig:rgz_misclassified_sources_3}}\\
\vspace{-0.2cm}
\subtable[\scriptsize{\texttt{2C-2P} (true), \texttt{1C-3P} (pred), \texttt{2C-3P} (true corr.)}]{\includegraphics[scale=0.25]{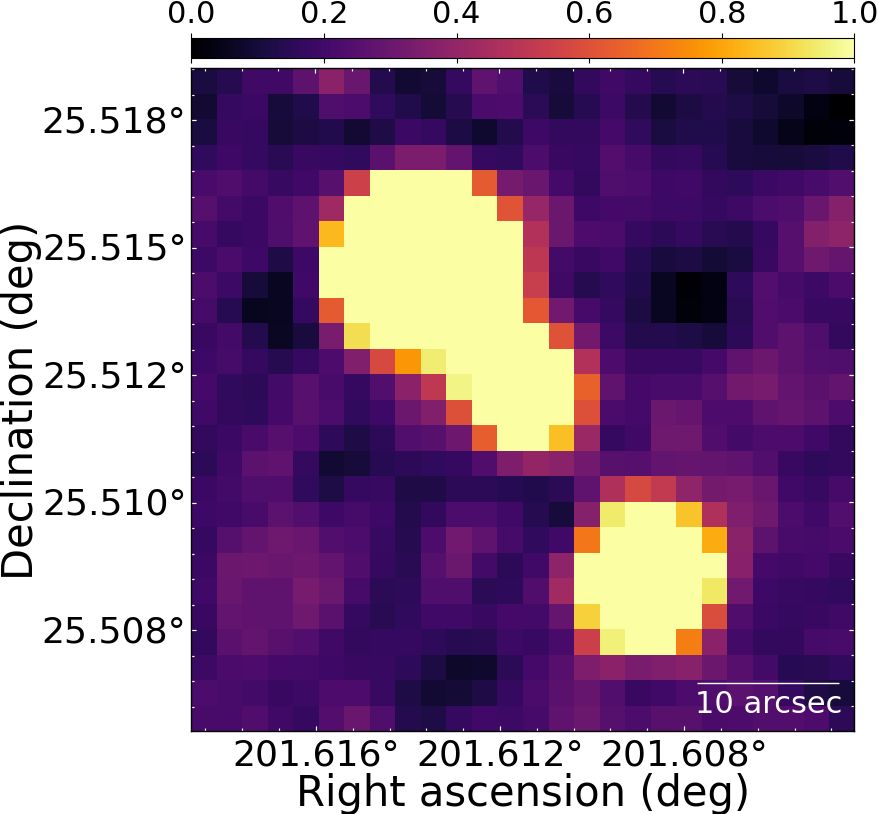}\label{fig:rgz_misclassified_sources_4}}
\subtable[\scriptsize{\texttt{2C-3P} (true), \texttt{3C-3P} (pred), \texttt{3C-3P} (true corr.)}]{\includegraphics[scale=0.25]{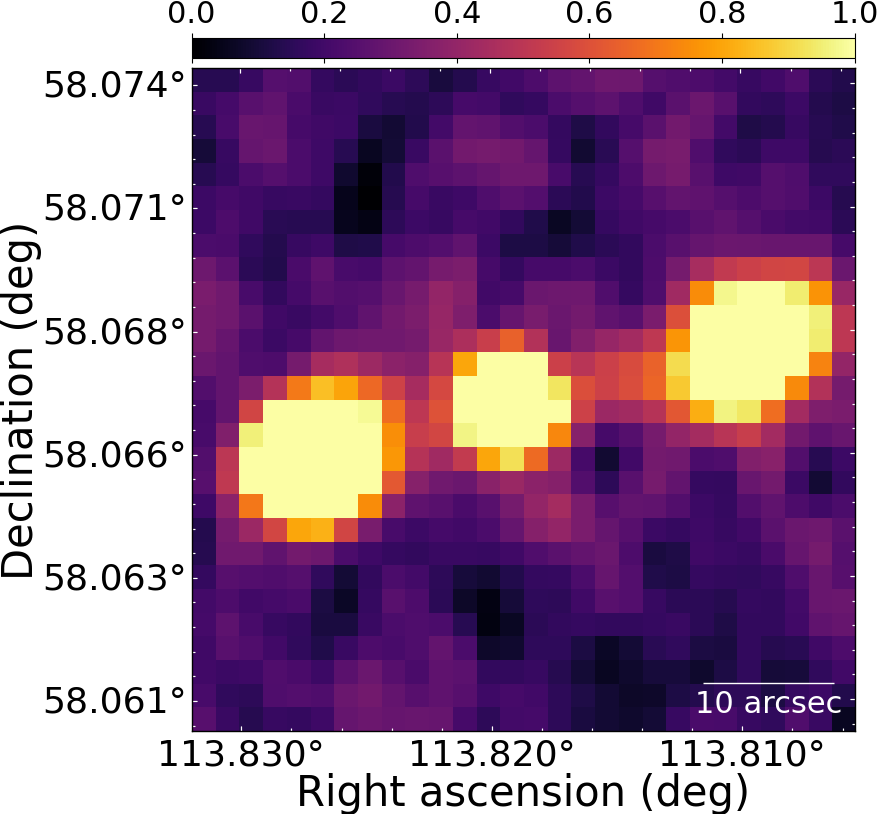}\label{fig:rgz_misclassified_sources_5}}
\subtable[\scriptsize{\texttt{3C-3P} (true), \texttt{2C-3P} (pred), \texttt{2C-3P} (true corr.)}]{\includegraphics[scale=0.25]{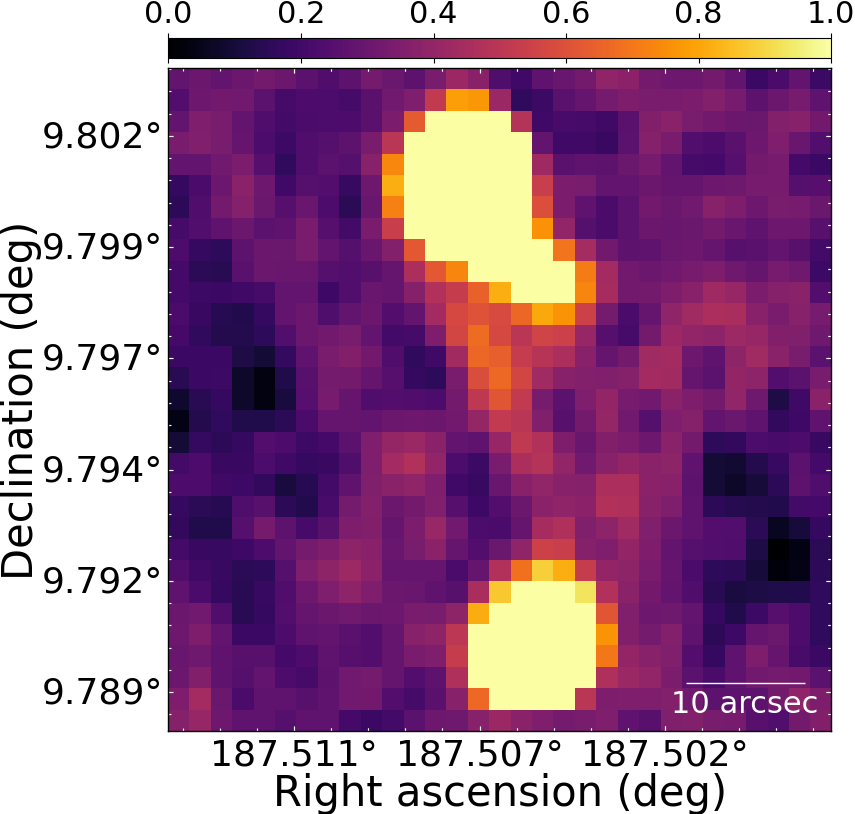}\label{fig:rgz_misclassified_sources_6}}
\caption{Examples of sources from the RGZ test dataset that were misclassified by the trained source classifier (\texttt{hulk\_smgps} pre-trained and frozen backbone) due to an incorrect true class label provided in the dataset (mislabelling). The true and predicted class labels are reported below each frame. In many cases, the model indeed correctly predicted the expected true classification (denoted as "true corr." below each frame).}%
\label{fig:rgz_misclassified_sources}
\end{figure*}

\begin{figure*}[htb]
\centering%
\includegraphics[scale=0.8]{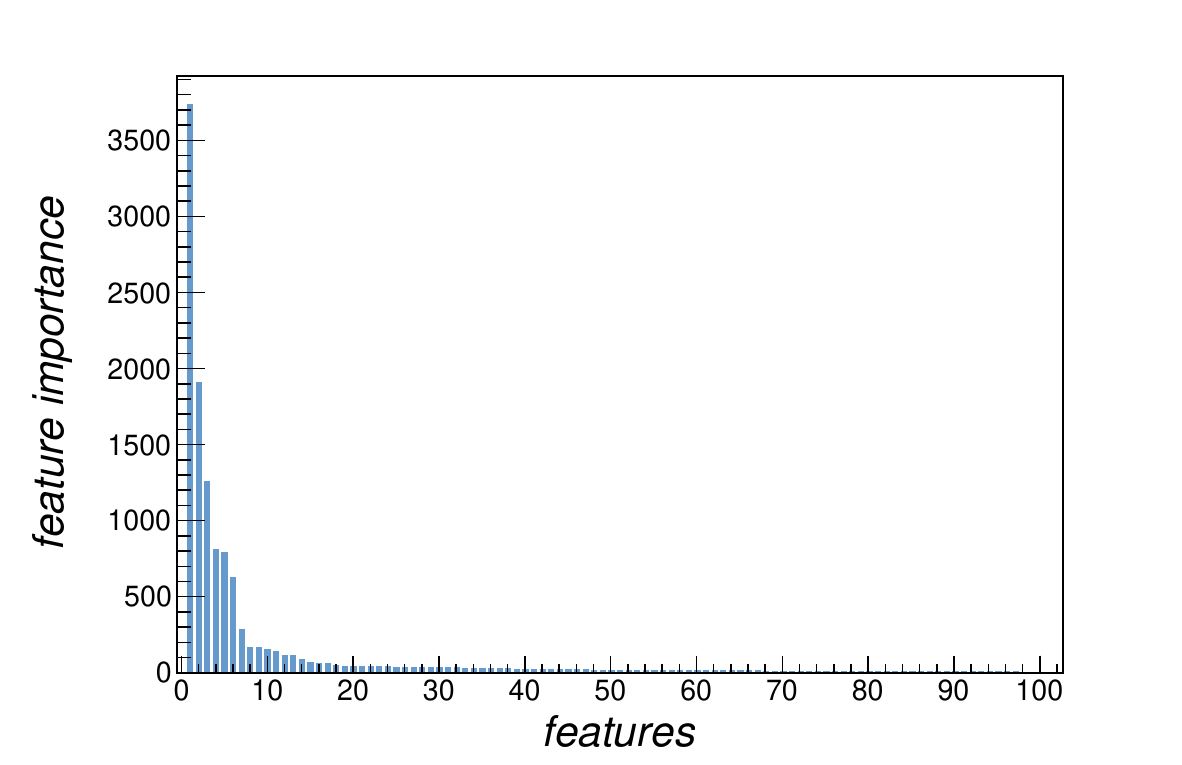}%
\caption{Feature importance obtained with a LightGBM classifier trained on \texttt{hulk\_emupilot} data representation, for the classification of interesting against ordinary images.}%
\label{fig:lgbm-feat-importance}
\end{figure*}

\begin{figure*}[htb]
\centering%
\subtable[Representation vs RMS]{\includegraphics[scale=0.27]{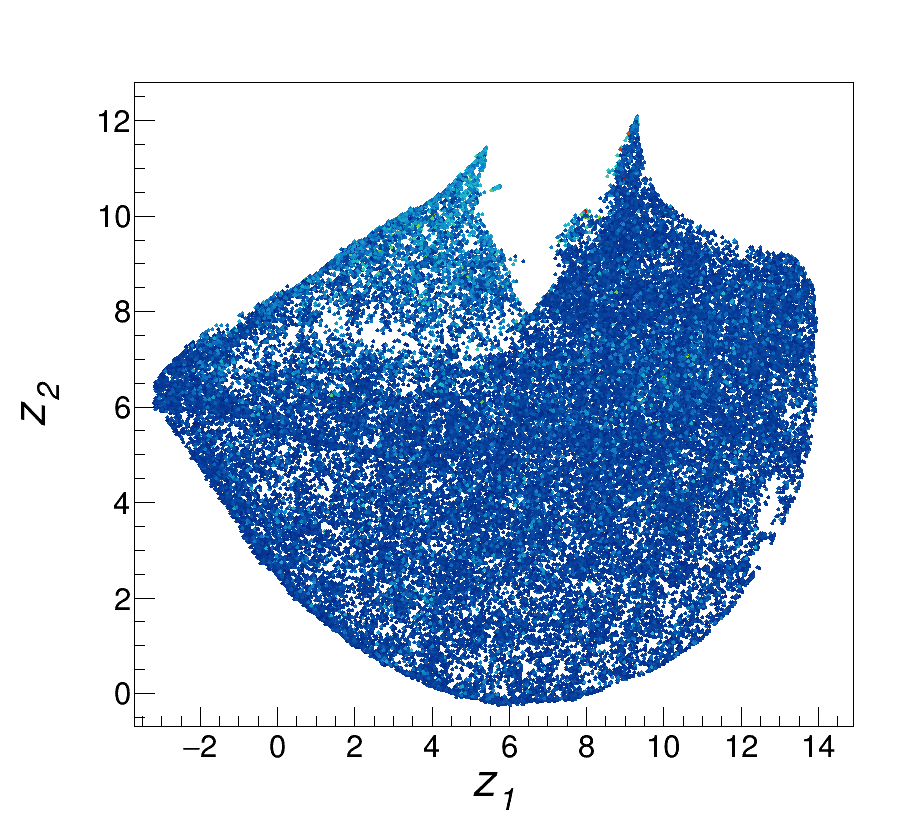}\label{fig:hulk_umap_vs_pars_1}}
\subtable[Ordinary frames]{\includegraphics[scale=0.27]{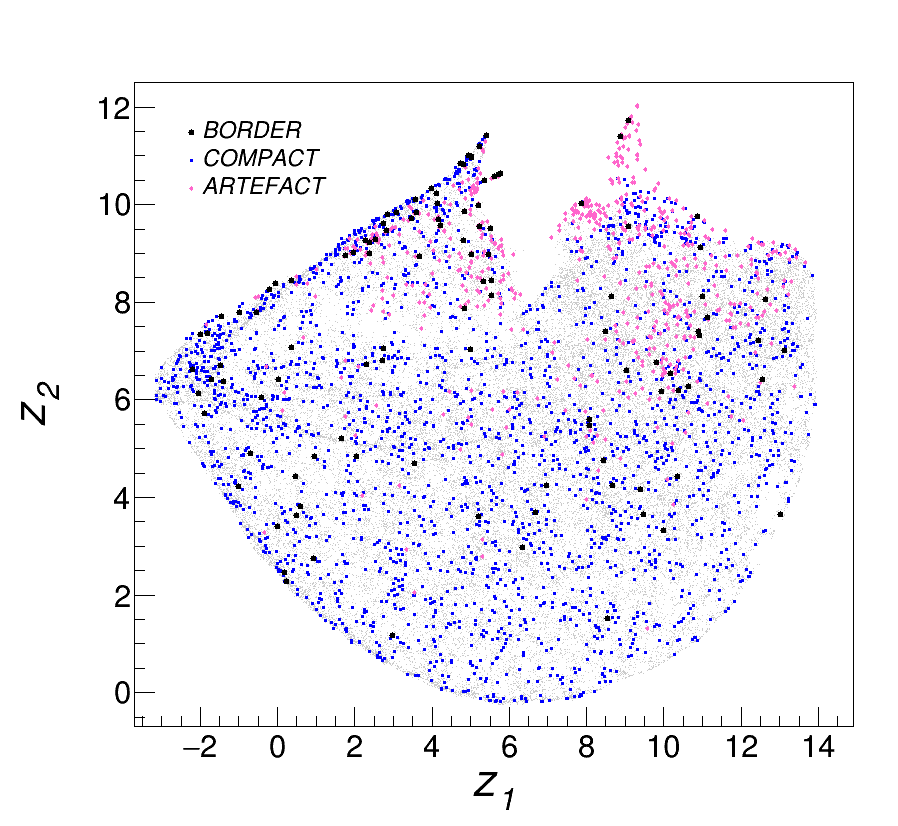}\label{fig:hulk_umap_vs_pars_2}}\\
\vspace{-0.4cm}%
\subtable[Complex frames]{\includegraphics[scale=0.27]{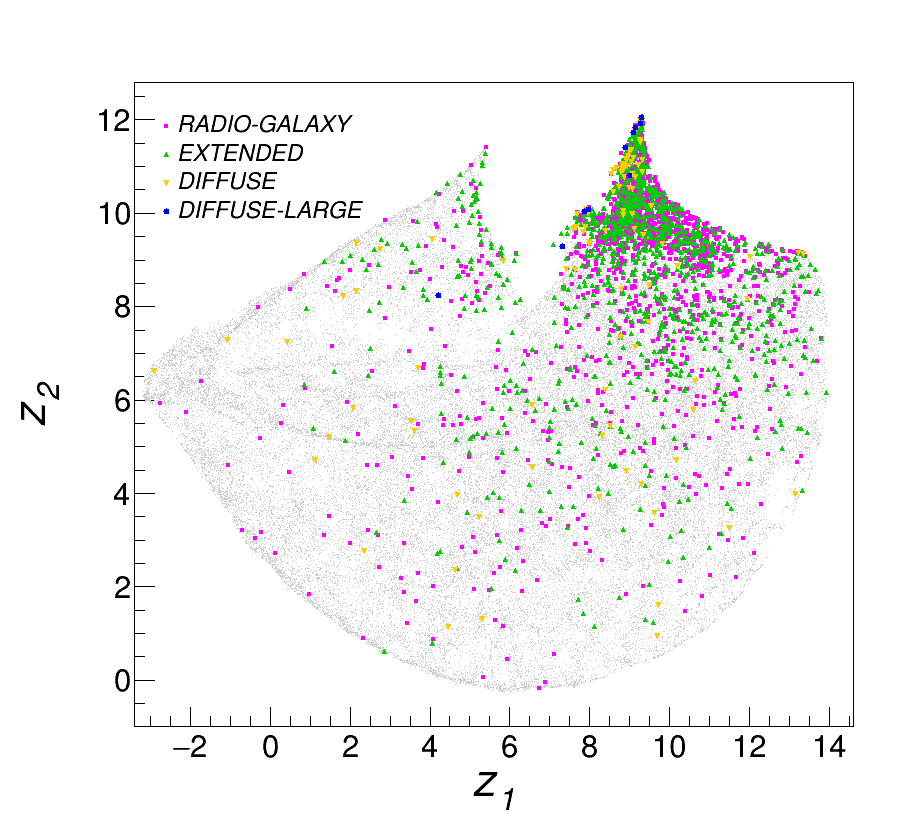}\label{fig:hulk_umap_vs_pars_3}}
\subtable[Peculiar frames]{\includegraphics[scale=0.27]{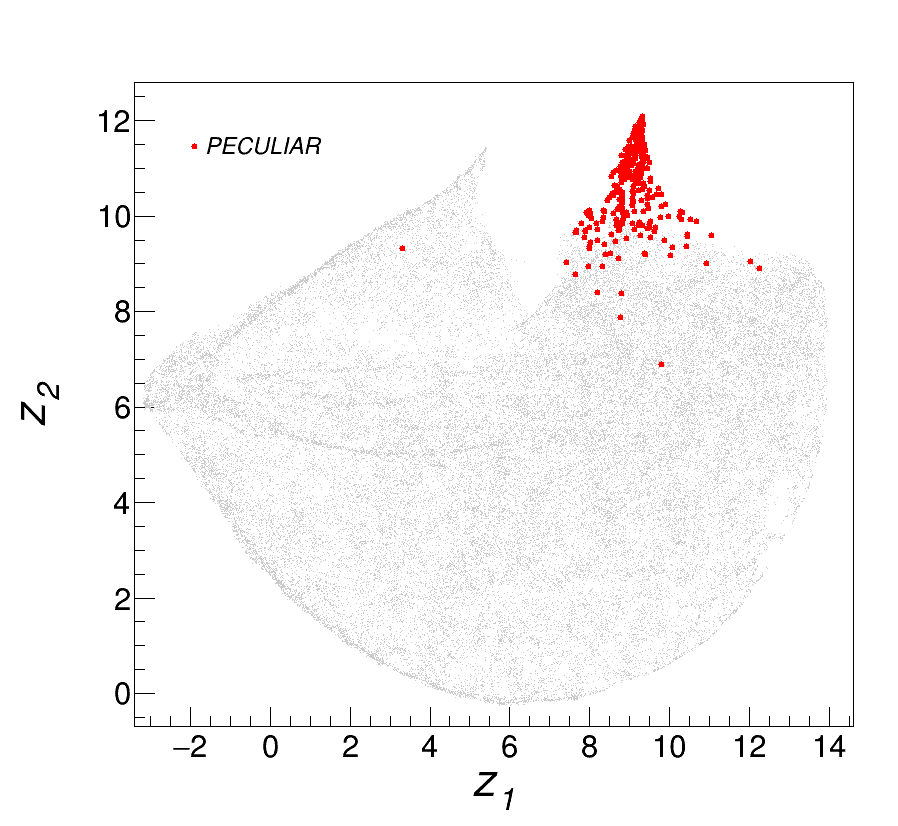}\label{fig:hulk_umap_vs_pars_4}}
\caption{Fig.~\ref{fig:hulk_umap_vs_pars_1}: 2D UMAP projection of the top-15 selected features from the data representation vector produced by the trained SimCLR model on the \texttt{hulk\_emupilot} dataset as a function of the image noise RMS level in logarithmic scale (z-scale axis). Red markers correspond to image with higher RMS levels, while blue markers to low noise RMS levels. Left: 2D UMAP projection of the top-15 selected features for unclassified frames (gray markers) and labelled frames (coloured markers, as reported in the plot legends). See text for details on label schema.}%
\label{fig:hulk_umap_vs_pars}
\end{figure*}

\begin{figure*}[htb]
\centering%
\includegraphics[scale=1.0]{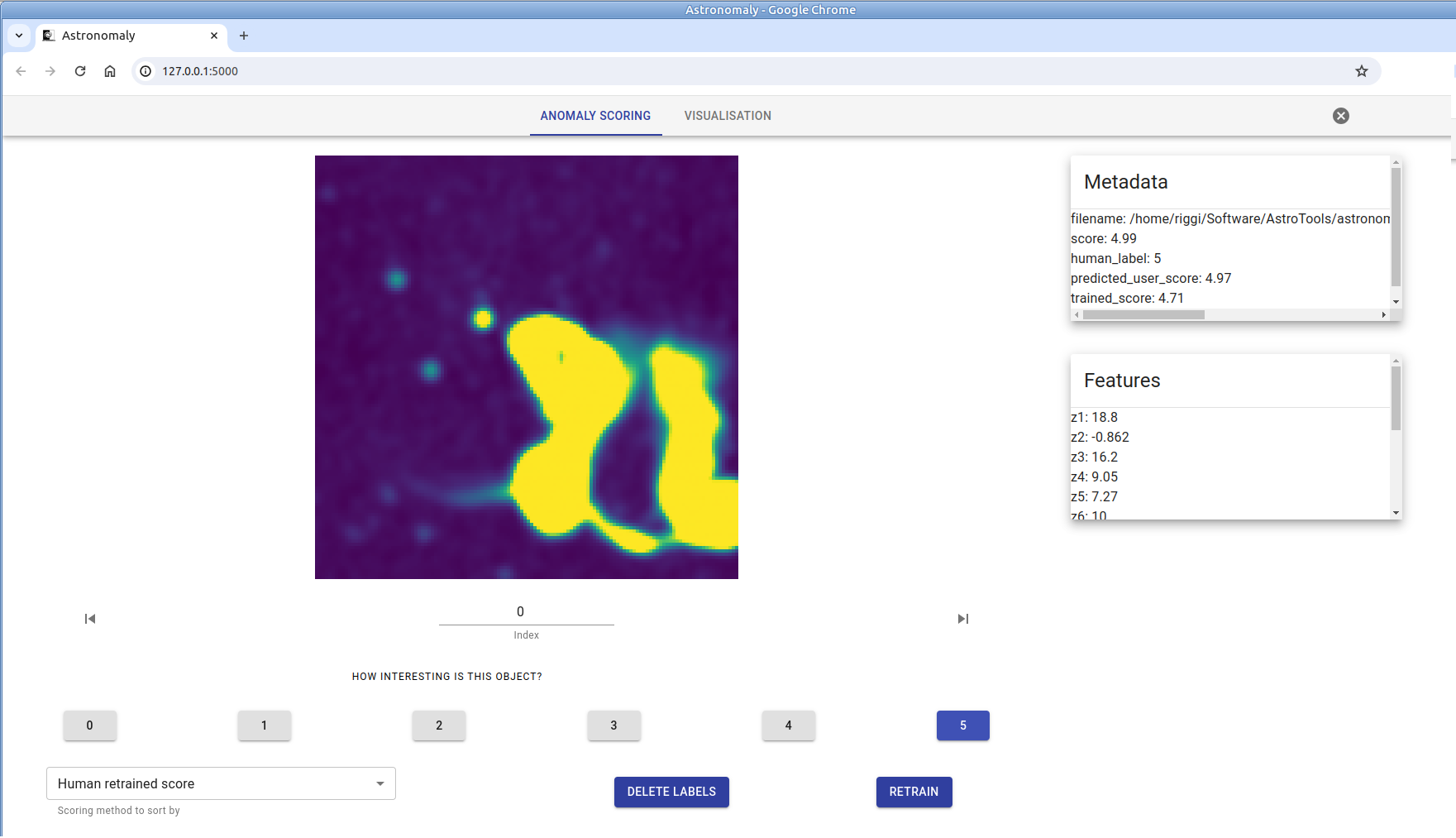}%
\caption{Screenshot of \texttt{astronomaly} web UI with list of anomalous frames selected from the \texttt{hulk\_emupilot} dataset.}%
\label{fig:astronomaly}
\end{figure*}

\end{document}